 \documentclass[final,5p,times,twocolumn,authoryear]{elsarticle}

\usepackage{amssymb}
\usepackage{amsmath}
\usepackage{lipsum}
\usepackage{xcolor}

\usepackage[normalem]{ulem} 
\newcommand{\considerdelete}[1]{%
  \bgroup 
  \renewcommand{\ULthickness}{0.1pt}
  \textcolor{lightgray}{\sout{#1}}
  \egroup 
}

\usepackage{booktabs}
\usepackage{array}
\usepackage{longtable}
\usepackage{url}

\usepackage{hyperref}
\usepackage{xurl}
\hypersetup{hidelinks} 

\journal{Astronomy $\&$ Computing}

\begin{document}

\begin{frontmatter}

\title{Variability in Cosmological Hydrodynamical Simulations: how Stochastic Processes, Numerical Effects, and Reproducibility Limits impact Predictability}

\author[INAF,UNITS]{Chaitra}
\affiliation[INAF]{organization={INAF (Italian National Institute for Astrophysics) – Astronomical Observatory of Trieste},
            city={Trieste},
            postcode={34143}, 
            country={Italy}}
\affiliation[UNITS]{organization={Department of Physics, University of Trieste},
            city={Trieste},
            postcode={34143}, 
            country={Italy}}
\affiliation[INFN]{organization={Instituto Nazionale di Fisica Nucleare},
            city={Trieste},
            postcode={34127}, 
            country={Italy}}
\affiliation[ICSC]{organization={Italian Research Center on High Performance Computing, Big Data and Quantum Computing},
            city={Casalecchio di Reno},
            postcode={40033}, 
            country={Italy}}
\affiliation[IFPU]{organization={Institute for Fundamental Physics of the Universe},
            city={Trieste},
            postcode={34151}, 
            country={Italy}}
\author[INAF]{Antonio Ragagnin}
\author[UNITS,INAF,INFN,ICSC]{Milena Valentini}
\author[INAF,IFPU,ICSC]{Giuseppe Murante}
\author[UNITS,INAF,INFN,IFPU,ICSC]{Stefano Borgani}
\author[INAF,ICSC,UNITS]{Giuliano Taffoni}

\begin{abstract}
Cosmological hydrodynamical simulations are powerful tools for studying galaxy formation, yet their predictive precision is limited by stochastic variability and numerical uncertainty. We quantify this variability using four identical realizations of a zoom-in galaxy-cluster simulation evolved with \textsc{OpenGadget3} under tightly controlled compiler, library, and hardware settings. Variability is measured through the properties of matched galaxies across repeated runs, including a mixed linear model that separates run-to-run variation from within-run noise.

Variations of approximately $10$–$25\%$ are found in galaxy dark matter and stellar masses for the baseline simulations. The variability trending above the shot-noise floor reflects the combined effects of stochastic star formation and feedback regulation, and is further amplified when black hole physics is included.

Furthermore, our results indicate that feedback acts to regulate variability, reducing scatter in both stellar and black hole masses. Our inference from run-to-run variation indicates a noise-dominated regime that remains statistically reproducible, despite individual realization differences.

These results establish baseline, noise-dominated variability estimates at low resolution, demonstrate how feedback modulates predictability, and provide a statistical framework for future studies of reproducibility in cosmological hydrodynamical simulations.
\end{abstract}


\end{frontmatter}

\section{Introduction}
\label{introduction}

Cosmological hydrodynamical simulations have become essential tools for understanding galaxy formation and evolution, enabling researchers to explore complex physical processes that span vast ranges in time and scale. Major simulation projects like Illustris, EAGLE, and IllustrisTNG \citep{vogelsberger_introducing_2014, schaye_eagle_2015, pillepich_simulating_2018} have provided crucial insights into how galaxies form within the cosmic web, how feedback processes regulate star formation, and how black holes coevolve with their host galaxies.

However, a fundamental challenge emerges when using these simulations to make quantitative predictions: their outcomes exhibit inherent variability that can significantly affect scientific conclusions. Even when simulations are initialized with seemingly identical conditions, individual galaxy properties can vary substantially between runs, raising critical questions about the reliability and interpretation of simulation predictions (see e.g. \cite{thiebaut_onset_2008, su_discrete_2018, keller_chaos_2019, genel_quantification_2019, davies_quenching_2021,davies_galaxy_2022,davies_are_2023, borrow_impact_2023}).

This variability in cosmological simulations arises from multiple distinct sources. At the most fundamental level, gravitational N-body systems can exhibit sensitive dependence on initial conditions (a hallmark of chaotic behavior), as first recognized in astrophysical contexts by e.g. \cite{miller_irreversibility_1964}, \cite{standish_numerical_1968}. However, modern cosmological simulations incorporate additional complexity through stochastic subgrid physics models for star formation, stellar feedback, and black hole accretion, each introducing algorithmic stochasticity that can amplify small differences into large-scale variations in galaxy properties.

Variability in cosmological simulations has been investigated in several studies using different simulation codes (e.g. AREPO, SWIFT, GASOLINE2, RAMSES). While small-scale perturbations are expected to average out in ensemble statistics of phase-space quantities, individual galaxy properties can remain sensitive to such perturbations. These works have reported repeated-run variations of approximately $2$–$25\%$ in various galaxy properties (e.g. \cite{genel_quantification_2019, keller_chaos_2019, borrow_impact_2023, pakmor_quantifying_2025}). The magnitude of the variation depends on both resolution and subgrid physics, with more stable quantities (e.g. halo mass, $M_\mathrm{dm}$, $V_\mathrm{max}$) showing smaller scatter, while higher variability is associated with processes such as mergers, stochastic or "bursty" feedback, and black hole growth and dynamics (In \ref{sec:appendix_variability_sources}, we provide a brief overview of the numerical and physical mechanisms that contribute to variability in cosmological simulations and summarize relevant findings from the literature).

Pure N-body simulations can be argued to be exhibiting chaotic behavior despite the presence of round-off error. But, when extending to hydrodynamic simulations incorporating subgrid models (i.e. for star formation, black hole etc.), any observed "chaotic-like" behavior is strongly influenced by stochastic processes. Thus, in cosmological simulations, the presence of stochasticity complicates or even prevents the unambiguous identification of effects of chaos, as well as knowing whether stochastic processes suppress or amplify underlying chaotic effects.

A more precise description would treat the observed divergence as a consequence of stochastic variability, possibly with features of stochastic-chaos, but not attributable to deterministic chaos in the classical sense.

Understanding the sources of simulation variability is crucial for several reasons. First, it determines the appropriate uncertainty estimates for simulation predictions, affecting how we compare simulations to observations. Second, misidentifying the dominant sources of unpredictability can lead to incorrect conclusions about the underlying physics of galaxy formation.

This study expands on earlier work in three key directions. First, it provides the first variability analysis performed using \textsc{OpenGadget3}, a widely used cosmological code for which stochastic divergence has not yet been quantified. Second, we introduce a controlled, module-based comparison that isolates the impact of individual physical processes such as stellar physics (cooling, star formation, stellar winds, and supernova (SN) feedback) and black hole physics (seeding and active galactic nucleus (AGN) feedback) on variability growth. This helps us address the question on whether feedback prescriptions amplify or regulate the growth of variations. Third, we apply a mixed linear modeling framework to statistically separate numerical noise from intrinsic run-to-run divergence, providing a structured way to assess convergence across repeated simulations.

The paper is structured as follows: We present the simulation setup and data processing in Section \ref{sec:methodology}, followed by the statistical methods for variability analysis in Section \ref{sec:statistical_analysis}. Section \ref{sec:results} presents the results for simulations without black holes, including both the baseline and feedback variation tests. Section \ref{sec:results_withbh} extends this analysis to simulations with black holes, comparing different feedback strengths and also discussing the effects observed in mixed precision configurations. Finally, Section \ref{sec:summaryConclusions} summarizes the main findings and outlines the implications for future simulation studies. 
Supplementary details are provided in the Appendices. These include results for dark-matter-only, dark-matter-plus-hydrodynamical simulations, tests using varied MPI task counts, and additional variability results for simulations with black holes and mixed precision configuration. Finally, we provide a technical description of the stochastic subgrid physics and random number generator (RNG) implementation.

\section{Methodology}
\label{sec:methodology}

Our aim is to quantify a baseline of total variability for cosmological hydrodynamical simulations using \textsc{OpenGadget3}. A key aspect of our methodology is the systematic isolation of astrophysical processes to determine their individual contributions to this divergence. We proceed progressively: first establishing baselines with stellar feedback alone, then incorporating black hole physics to assess their additional impact. This allows us to disentangle the influence of specific mechanisms rather than merely observing their combined effect, providing clearer insights into the drivers of simulation variability.

The experimental framework consists of four main components: (1) a suite of identical simulations to establish baseline variability under controlled conditions, (2) systematic variations in feedback parameters to isolate the effects of different algorithmic stochasticity, (3) a galaxy matching procedure to track individual objects across simulation realizations, and (4) statistical methods to decompose total variability into its constituent sources. 

\subsection{\textsc{OpenGadget3}: Code and Physical Modules}
All simulations in this study were conducted using \textsc{OpenGadget3}, a novel implementation of the Gadget code \citep{springel_gadget_2001}.
We employ a stable version of \textsc{OpenGadget3} (consistent with \cite{groth_cosmological_2023} and configured similarly to \cite{bassini_dianoga_2020}) that provides reliable performance and flexibility for extensive parameter exploration. The code uses TreePM for gravitational forces and Smoothed Particle Hydrodynamics (SPH) for gas dynamics, with adaptive time stepping to handle the vast range of dynamical timescales in cosmological simulations\footnote{
For reference, the project’s repository is: \url{https://www.space-coe.eu/codes/opengadget.php}. The specific production version used in this work is hosted at: \url{https://www.ict.inaf.it/gitlab/tscosmo/opengadget3-triesteproductionversion}. At the time of publishing of this paper, OpenGadget3 is a private code with its community-wide release planned in the forthcoming months; we provide these repositories for reference until the code base is available publicly. The code will also be shared on request.}.

The simulation code incorporates comprehensive subgrid physics following the implementations detailed in \cite{bassini_dianoga_2020}. Key physical processes include:

\begin{itemize}
    \item Cooling and metal enrichment \citep{tornatore_cooling_2003, tornatore_chemical_2007}
    \item Star formation following \cite{springel_cosmological_2003} with Schmidt law-like prescriptions
    \item Stellar feedback including supernova explosions, galactic winds, and chemical enrichment
    \item Black hole physics \citep{springel_modelling_2005} with Bondi-Hoyle-Lyttleton accretion and thermal feedback (when included)
\end{itemize}

A summary of the stochastic subgrid physics and RNG implementation used in this study is provided in \ref{sec:appendix_intro_sf}. For our controlled experiments, we systematically enable and disable specific components to isolate their individual contributions to simulation variability.

\begin{table*}[h!]
    \centering
    \caption{Simulation configuration and cosmological parameters}
    \begin{tabular}{c|c|c|c}
        \toprule
        \multicolumn{4}{c}{\textbf{Dianoga D1-run configuration }} \\
        \midrule
        \textbf{Resolution} & \(m_{\text{DM}}\) & \(m_{\text{gas,initial}}\) & \(\epsilon_{\text{DM}}\) \\
        \midrule
        1x & 8.43 \(\times 10^{8} M_\odot h^{-1}\) & 1.56 \(\times 10^{8} M_\odot h^{-1}\) &  \begin{tabular}{@{}c@{}}5.62 kpc \(h^{-1}\) \\ for \(z > 2\): 16.87 kpc \(h^{-1}\)\end{tabular} \\
        \midrule
        \multicolumn{4}{c}{\textbf{$\Lambda$CDM Model Parameters}} \\
        \midrule
        \(\Omega_M\) & \(\Omega_B\) & \(h\) & \(\sigma_8\) \\
        \midrule
        0.24 & 0.04 & 0.72 & 0.8 \\
        \bottomrule
    \end{tabular}
    \label{tab:simconfig}
\end{table*}

\subsubsection{Initial Conditions}
The  zoom-in initial conditions used in this project for all simulations are the ones described  by \cite{bonafede_non-ideal_2011}.
They are the so-called "Dianoga-set" of simulations (see \cite{ragone-figueroa_brightest_2013}, \cite{rasia_cool_2015}, \cite{planelles_pressure_2017}, \cite{biffi_history_2017}, \cite{biffi_origin_2018}, \cite{ragone-figueroa_bcg_2018}, \cite{bassini_black_2019} and \cite{bassini_dianoga_2020}) that were generated from a parent low-resolution dark matter only simulation of comoving cubic volume 1 (Gpc \(h^{-1}\))\textsuperscript{3} executed with GADGET-2 code and adopting a $\Lambda$CDM 
cosmological model (see Table \ref{tab:simconfig}). The dark matter particle has a mass of approximately $10^9 \, h^{-1} M_\odot$. 
This provides a relatively coarse initial resolution, sufficient to identify and track 
the formation of large-scale structures, including massive galaxy clusters. 

The box is then processed with Friends-of-Friends (FOF; \cite{davis_evolution_1985}) to identify the most massive clusters: 24 in number, with \(M_{200} > 8 \times10^{14}h^{-1}M_{\odot}\) and 5 relatively less massive with \(M_{200}\) ranging from 1-4 \(\times 10^{14} h^{-1} M_{\odot}\)). 

The volumes selected for each object are then simulated at different (higher) resolutions 
with zoomed initial condition technique (ZIC code, see \cite{tormen_structure_1997} and \cite{bonafede_non-ideal_2011}). 
Each initial condition (IC) for the selected objects was generated by tracing the position of all particles within this Lagrangian region (a sphere of 5-7 virial radius at  $z = 0$ centered on the respective object) all the way up to the beginning. The region traced in the parent box is divided into \( 64^3 \) cells, and the cells are resampled with higher resolution collisionless particles. 

For simulations with full physics (gas, star formation, black holes etc.), baryons are added to the ICs along with dark matter. 
Thus ICs of different resolutions are available for running both dark matter only and full physics simulations.

For our work, we focus on one galaxy cluster (D1/g0016649) with virial mass $M_{\rm vir}= 1.6\times10^{15} M_\odot$, simulated at the lowest resolution level (Dianoga 1x). Dark matter particle mass is $8.43 \times 10^8 M_\odot h^{-1}$, initial gas particle mass is $1.56 \times 10^8 M_\odot h^{-1}$, and gravitational softening lengths are 5.62 kpc $h^{-1}$ for gas/dark matter and 3.0 kpc $h^{-1}$ for stars/black holes.

\subsubsection{Computational Controls}
In studies of simulation variability, it is critical to tightly constrain numerical factors because even subtle variations in code setup, compiler optimizations, libraries, or hardware architecture can affect results. Different code versions may contain algorithmic changes, bug fixes, or optimizations that alter operation ordering or numerical precision handling.

\textit{Library and Compiler}: We fix all library versions (FFTW 3.3.10, GSL 2.7.1, HDF5 1.12.2) and use GCC 11.2.0 across all simulations to ensure reproducible floating-point operations. 
The exact compiler flags which fix the C++ standard and optimization levels, used for all simulations were: -std=c++11 -Wall -g -O2

We adopt OpenMPI 4.1.3 and maintain consistent parallelization schemes to avoid variations in operation ordering that can arise from different MPI implementations. Processes are not bound to specific cores or sockets (with the --bind-to none flag) and OpenMP was inactive.

\textit{Hardware and Execution Environment}: All simulations run on the Hotcat cluster \citep{taffoni_chipp_2020, bertocco_inaf_2020} using identical Intel Xeon CPU E5-2697v4 processors (2.30GHz) with 256GB memory. We maintain consistent computational resources: 4 computing nodes with 36 cores each, using 1 MPI task per core (144 total cores). Identical executables are used for all runs except when feedback parameter modifications require recompilation.

\subsubsection{Parameter Variations}
Our experimental design systematically varies feedback parameters while maintaining all other conditions identical, enabling isolation of specific physical process contributions to simulation variability. For each parameter configuration, we run four identical simulations to quantify repeated-run variability. This sample size balances computational feasibility (each simulation test-set requires a total of \(\approx 1.65 \times 10^4\) CPU core-hours, including postprocessing and analysis) with adequate sampling to characterize variability.

\textbf{The Fiducial Model}: The fiducial configuration incorporates comprehensive hydrodynamics and standard subgrid physics. For stellar feedback, the fiducial wind velocity is $v_w = 350$ km s$^{-1}$, corresponding to approximately 50\% of supernova energy converted into kinetic energy carried by galactic winds \citep{barai_galactic_2015, bassini_buchi_2021}.

\begin{table}[h!]
    \centering
    \caption{Feedback parameter configurations tested}
    \begin{tabular}{l|c|c}
        \toprule
        \textbf{Configuration} & \textbf{Stellar Feedback} & \textbf{Black Hole Feedback} \\
         & $v_w$ (km s$^{-1}$) & $\epsilon_r$ \\
        \midrule
        \multicolumn{3}{c}{\textit{Phase 1: Stellar feedback only (no black holes)}} \\
        \midrule
        FIDUCIAL & 350 & N/A \\
        LOW & 200 & N/A \\
        HIGH & 500 & N/A \\
        ZERO & 0 & N/A \\
        \midrule
        \multicolumn{3}{c}{\textit{Phase 2: Combined stellar and black hole feedback}} \\
        \midrule
        FIDUCIAL & 350 & 0.1 \\
        HIGH & 500 & 0.2 \\
        ZERO & 0 & 0.0001 \\
        \bottomrule
    \end{tabular}
    \label{tab:feedback_configs}
\end{table}

\textbf{Stellar Feedback Parameter Space}: To isolate stellar feedback effects, we first conduct experiments without black hole physics by setting the critical halo mass for black hole seeding to an unphysical high value. The parameter ranges (Table~\ref{tab:feedback_configs}) span physically motivated values consistent with previous cosmological simulations (e.g. \cite{barai_galactic_2015, bassini_buchi_2021}).

\textbf{Black Hole Feedback Parameter Space}: In the second phase, we introduce black hole physics while maintaining stellar feedback configurations. The radiative efficiency values encompass the range commonly used in cosmological simulations (e.g. \cite{springel_modelling_2005}).

\begin{figure*}[htbp]
    \centering
    \includegraphics[width=0.49\textwidth]{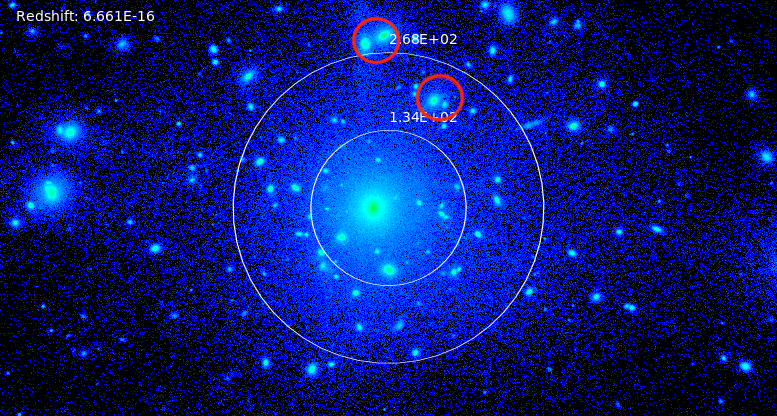}
    \includegraphics[width=0.49\textwidth]{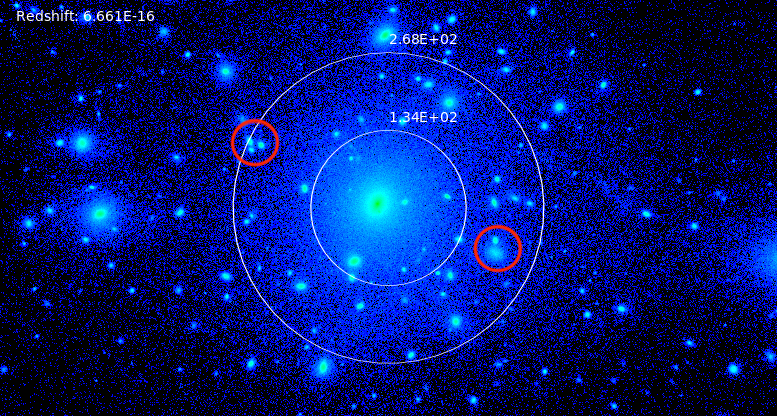}
    \includegraphics[width=0.49\textwidth]{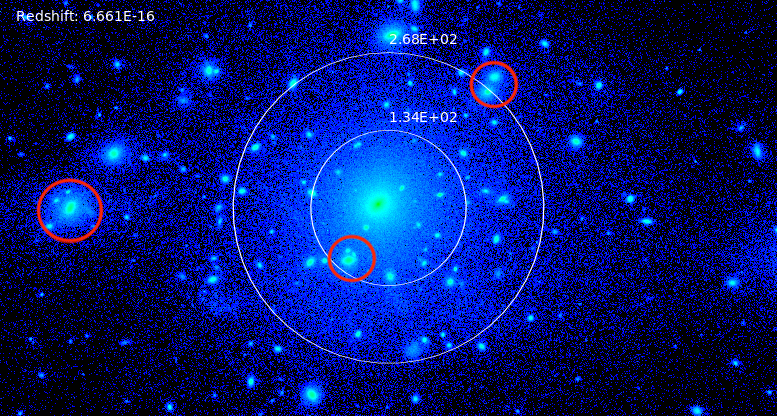}
    \includegraphics[width=0.49\textwidth]{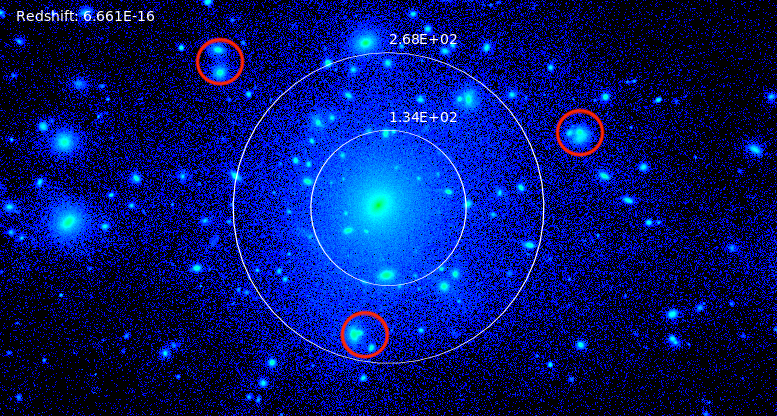}
    \caption{Visual representation of the Dianoga simulation for the Fiducial model showing 4 realizations of the same galaxy cluster at $z=0$. Substructure differences are encircled in red.}
    \label{fig:highlighted_D1_fiducial_4N}
\end{figure*}

\subsection{Galaxy Identification and Matching}
\label{subsec:galaxy_identification}

This study quantifies variability through changes in macroscopic galaxy properties rather than individual particle trajectory divergence, capturing the practical impact of simulation unpredictability on observable cosmological phenomena. In practice, this approach tracks matched "clone" galaxies across simulation runs using unique particle IDs.

Figure \ref{fig:highlighted_D1_fiducial_4N} provides a visual demonstration of the variability explored in this work, showing the same galaxy cluster at $z=0$ across four identical runs of our Fiducial model (for comparison, analogous visualizations for dark-matter-only and hydro-only configurations are provided in \ref{sec:appendix_dmo}).
Although the central galaxy remains morphologically similar across realizations, measurable differences are present. We focus our analysis on the Brightest Cluster Galaxies (BCGs) and their progenitors at high redshifts. These galaxies, which are the central subhalos of their respective halos, are chosen for several methodological reasons: they represent the most massive and luminous galaxies in each cluster and are better resolved in our low-resolution simulations, allowing for reduced shot noise.
In contrast, the visible differences in satellite positions and structures across runs (highlighted in Figure 1) are apparent. Satellites are particularly sensitive to small perturbations in timing, merger histories, and environmental interactions.  Thus our approach deliberately limits the analysis to objects expected to show lower scatter, although it may underestimate total system variability.

The zoom-in region of the D1 cluster contains a large catalog of halos. We analyze the central subhalo of every identified halo within this volume, obtaining approximately 450-600 galaxies per snapshot (see Section \ref{subsec:data_prep}). Galaxy identification is achieved through post-processing using FOF and SUBFIND algorithms \citep{springel_e_2010, dolag_substructures_2009}. Starting from one simulation as reference, we identify matching galaxies by selecting the most massive central galaxy within each parent halo across identical runs. For this work we use g3read and g3matcha \citep{ragagnin_aragagning3read_2025}.

Our matching procedure involves two steps: (1) identify clone halos by requiring $>60\%$ overlap in particle IDs and $>50\%$ mass agreement, and (2) apply similar criteria to central galaxies within matched halos. We impose a minimum mass threshold of $10^{11} M_\odot$ to ensure robust comparisons. These thresholds were chosen to balance completeness with accuracy based on preliminary tests. 

While comprehensive matching methods such as \cite{angulo_one_2010}, \cite{schaller_baryon_2015}, \cite{lovell_fraction_2018}, \cite{bose_no_2019}, \cite{genel_quantification_2019} and \cite{borrow_impact_2023} can accommodate complex galaxy evolutionary processes (including merger events), this adapted approach provides a practical balance between accuracy and computational efficiency (considering the large catalog of galaxies in the D1 cluster). \(95\% - 98\%\) of galaxies are matched with the adapted technique; a higher refinement in the matching algorithm will be left for future work, particularly for the study of impacts of mergers which is not in the scope of current work.
For the matched galaxies, we extract properties from both SUBFIND catalogs (dark matter mass, stellar mass, black hole mass) and snapshot files (cold gas mass, additional stellar mass measurements using 50 physical kpc apertures, and black hole mass of the most massive black hole within a 30 pkpc aperture from the galaxy center).

\section{Statistical Analysis Framework}
\label{sec:statistical_analysis}

Following the simulation execution and galaxy matching procedures described in Section~\ref{sec:methodology}, we obtain catalogs of matched clone galaxies with measured properties across identical simulation runs. We conduct statistical analysis for each set of simulations, following a methodology similar to that employed by \cite{genel_quantification_2019}, where the variation is computed for a sample of galaxies matched across identical runs. This section presents our statistical framework for quantifying simulation variability. 

\subsection{Data Preparation and Preprocessing}
\label{subsec:data_prep}

Our dataset consists of matched galaxy catalogs from four identical simulations for each parameter configuration. The number of matched galaxies varies with redshift, averaging approximately 500 per snapshot across all runs. This provides a sufficiently large sample for statistical analyses, including the examination of variability for specific samples (e.g. variability for different halo mass bins). For cases where the number of matched galaxies falls below 50 at a given redshift, we do not report results to avoid biases from poor sampling. As expected, less regulated feedback models (ZERO and LOW) yield larger numbers of substructures, resulting in a higher average of matched galaxies ($\approx $600), whereas more efficient feedback runs (FID and HIGH) produce fewer systems, with averages of $\approx $450–550. Nonetheless, provided the sample size exceeds the adopted threshold, central tendencies remain reliable for comparative analysis across feedback configurations.

\textbf{Normalization}: To ensure consistent measurements across simulations, galaxy properties are normalized using the within-subject mean (i.e. matched-galaxy mean), thus accounting for differences in scaling between galaxies while preserving relative differences between the runs. This approach isolates repeated-run variations from intrinsic galaxy-to-galaxy differences. The data are log-transformed (using base-10) to stabilize variance and improve the approximation to normality required for statistical modeling.

Values with z-scores exceeding ±10 are excluded. This threshold removes only a small number of extreme outliers (typically 1–2 isolated points) that could disproportionately inflate the results. However, cases with numerous or systematically large deviations, which may reflect physically meaningful differences (e.g. due to varying feedback mechanisms), are retained deliberately.

\textbf{Galaxy Properties Analyzed}: We compute variation for galaxy properties including dark matter mass ($M_{\rm dm}$) and stellar mass ($M_*$) available from SUBFIND catalogs, as well as cold gas mass extracted from 50 physical kpc apertures centered on matched galaxies from snapshot files. This size is a commonly adopted choice in \textsc{OpenGadget}-based studies that captures the central galaxy while minimizing intracluster light (ICL) contamination and allowing for comparison with observational data \citep{ragone-figueroa_bcg_2018, bassini_buchi_2021, marini_velocity_2021} (although stellar mass variations computed for quantities extracted from 50 kpc apertures compared to those of the SUBFIND catalog were not found to be markedly different).

\subsection{Method 1: Pooled Standard Deviation}
\label{subsec:pooled_method}

Our primary approach computes pooled standard deviations within bins of time rather than at individual simulation snapshots. \(\Delta t=0.2\) in lookback time was adopted after testing for different bin sizes with an average of 40 bins depending on the test case. Binning mitigates noise from low-number statistics at individual snapshots while preserving temporal evolution of variability. 

The pooled standard deviation for each time bin is computed as:
\begin{equation}
\sigma_{\mathrm{pooled}} = \sqrt{ \frac{ \sum_{i=1}^{N} (n_i - 1)\sigma_i^2 }{ \sum_{i=1}^{N} (n_i - 1) } }
\label{eq:pooled_std}
\end{equation}

where $\sigma_i$ is the standard deviation (in log10 space) across identical runs for galaxy $i$, and $n_i$ is the number of available runs for that galaxy. 

\subsection{Method 2: Mixed Linear Model Decomposition}
\label{subsec:mlm_method}

The variation in galaxy properties is affected not just by run-to-run variation but also by noise within individual simulations. While run-to-run variation reflects systematic differences between simulations that would otherwise be identical, within-simulation noise originates from stochastic processes and computational limitations. 

Method 2 employs Mixed Linear Models (MLM; \cite{gelman_data_2006}) to separate these components. MLMs are particularly suited for hierarchical data where observations are nested within groups: in our case, each galaxy is measured across different simulation realizations (repeated measures).

\textbf{Model Specification}: The model decomposes galaxy properties into two variance components:

\begin{equation}
y_{e,i} = u_e + v_{e,i}
\label{eq:mlm_model}
\end{equation}

where:
\begin{itemize}
    \item $y_{e,i}$ represents a given property for matched galaxy $i$ in simulation run $e$
    \item $u_e \sim N(0, \sigma_u^2)$ is the random effect capturing variation between simulation runs, representing differences in overall mean galaxy properties between identical simulations i.e. run-to-run variation.
    \item $v_{e,i} \sim N(0, \sigma_v^2)$ is the residual error capturing variation within each simulation run, i.e. total noise
\end{itemize}

The model takes as input the dependent variable (e.g. matched galaxy stellar mass; see subsection \ref{subsec:data_prep}) arranged by simulation run. Thus, each observation consists of a galaxy ID, its simulation run label (e $\in$ {1,2,3,4}), and its measured property value.

\textbf{Parameter Estimation}: For each galaxy property, the MLM is fitted using Restricted Maximum Likelihood Estimation (REML), where fixed effects (group means) are first estimated, then variance components are estimated based on residual data. The total variation is partitioned into run-to-run variation and residual noise: both $\sigma_u^2$ and $\sigma_v^2$ are estimated. Here, "group means" refer to the estimated mean property value within each simulation run (i.e. the mean across all galaxies in run $e$).

\textbf{Total Variance Decomposition}: The total variability is expressed as:
\begin{equation}
\sigma_{\mathrm{total}}^2 = \sigma_u^2 + \sigma_v^2
\end{equation}

This decomposition enables assessment of relative contributions from systematic differences between identical simulations versus random noise within individual runs.

Only results from converged models are reported, using the statsmodels package \citep{seabold_statsmodels_2010}.

\subsection{Shot Noise Estimation}
\label{subsec:shot_noise}

 As discussed in \ref{sec:appendix_variability_sources}, a baseline level of variability is expected from finite particle sampling i.e. shot noise. An approximate lower limit for shot noise in measurements of galaxy dark matter mass and stellar mass is given by:

\begin{equation}
\sigma_{\mathrm{shot}} \approx \frac{1}{\sqrt{N}}
\end{equation}

where $N$ is the number of stellar or dark matter particles contributing to the galaxy property measurement. This serves as a lower limit on measurable total scatter, below which apparent variation cannot be meaningfully interpreted as physical or numerical divergence. Note, however, that individual components of the variability decomposition (e.g. the run-to-run variance from the MLM) may fall below this threshold, since the shot noise estimate constrains the total measurable variance, not its sub-components.

\subsection{Connection to Physical Interpretation}
\label{subsec:physical_connection}

A total $\sigma$ value thus computed for a given galaxy property with the described methods can be interpreted as a fractional scatter across otherwise identical realizations of the same system. In this sense, the results serve analogously to an "error bar" on the predicted quantity, representing the effective precision limit of the simulation.

Given the numerous sources of variability in cosmological simulations (as discussed in \ref{sec:appendix_variability_sources}), it remains challenging (if not impossible) to disentangle the contributions of numerical noise from those of run-to-run divergence to these "error bars". In this regard we employ a mixed linear model as an attempt to provide a statistical decomposition of the two. While the run-to-run or "chaotic-like" variability term may, in part, reflect divergence related to stochastic chaos, we adopt the neutral terminology "run-to-run variability" throughout for consistency. Note that in previous literature, the term "run-to-run variability" is often used to denote the total observed scatter, whereas here it specifically refers to the decomposed component of the total variability. We do not precisely quantify the level of each contribution at this stage, as such an analysis would require more complete tests including systematic resolution-based studies. Instead we focus on mainly comparative analysis between different simulation test sets, reporting only trends that are sufficiently evident rather than marginal changes. The second method is therefore intended as a framework for future investigations aimed at disentangling the relative roles of noise and "chaotic-like" divergence in driving variability.

\section{Variability without black holes}
\label{sec:results}
We begin by examining the simulations excluding black holes to establish a baseline measure of variability arising solely from stellar physics, feedback, and numerical noise. This provides a controlled reference against which the impact of black holes i.e. accretion, feedback and black hole dynamics (presented in the next section), can be directly assessed. By comparing the two sets, we can disentangle variability intrinsic to our simulation framework, specific subgrid physics modules, and their respective feedback processes.
\begin{figure}
    \centering
    \includegraphics[width=1\linewidth]{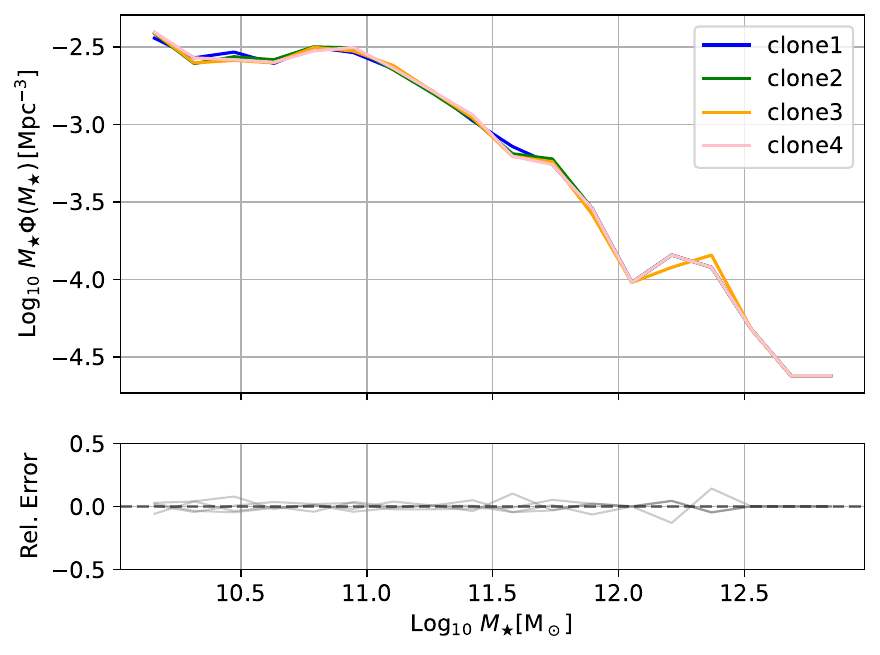}
    \caption{
Galaxy stellar mass function (GSMF) at \(z\approx0\) for Fiducial simulation set. Each colored line corresponds to an identical run (clone) in the set, considering aggregate of stellar mass reported by SUBFIND for all galaxies in the respective run. The lower panel shows the relative differences of each clone's GSMF with respect to the mean of the ensemble, illustrating the minimal clone-to-clone variation. The horizontal dashed line indicates zero deviation.}
    
    \label{fig:fidgsmfnobh}
\end{figure}

\begin{figure}
    \centering
    \includegraphics[width=1\linewidth]{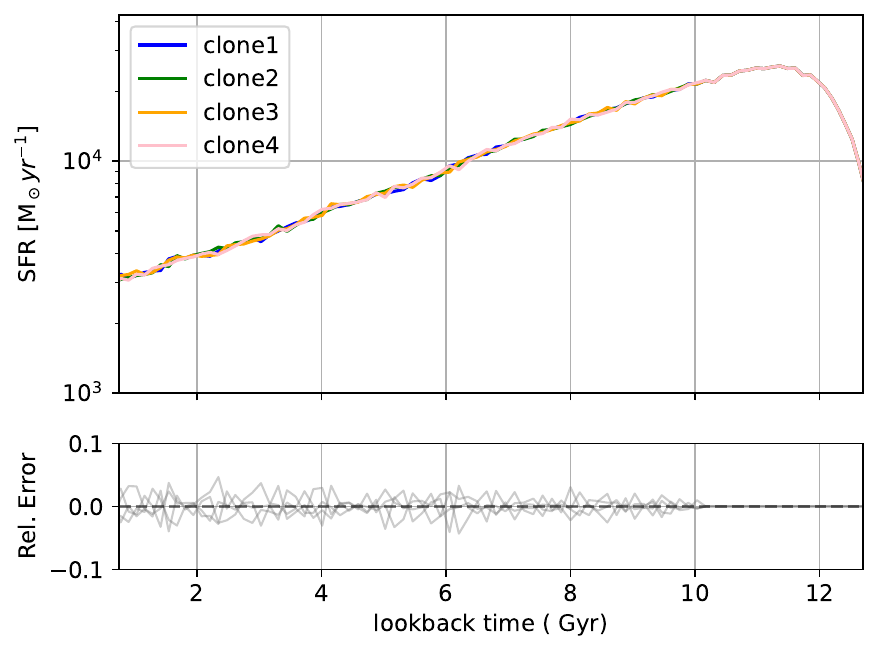}
    \caption{Global star formation history across the four clones of the Fiducial simulation set as a function of lookback time (top panel), where each colored line corresponds to an individual realization (clone). Bottom panel shows the relative variation of each clone's star formation rate compared to the ensemble mean.}
    \label{fig:fidsfrnobh}
\end{figure}

\subsection{Ensemble averaged properties in the simulation volume}
Galaxy stellar mass function (GSMF) for Fiducial simulation (see Table \ref{tab:feedback_configs}) set is shown in Figure \ref{fig:fidgsmfnobh}. The GSMF is constructed for all galaxies identified by SUBFIND at $z=0$, considering all star particles within 50 pkpc of a given galaxy center. The lower panel shows the GSMF from each individual realization (clone) expressed as a fractional difference from the ensemble mean. The values cluster tightly around zero across most of the mass range, indicating very small clone–to–clone variation. Figure \ref{fig:fidsfrnobh} shows the star formation history (SFH) constructed by following the evolution of star formation rate for the identical simulations in the Fiducial set. The star formation rate (SFR) presented here is not instantaneous SFR, but derived from the stellar ages of the particles and represents the global SFH of the simulation volume at $z=0$. The relative error of each realization (bottom panel) with respect to the mean SFR is minimal, signifying that the global SFH is a robust representation of the Fiducial model.

Together, these results provide a statistical overview across the population of galaxies in the four cloned simulations. The negligible discrepancies observed between the properties of galaxies among the clones, across a large population and on large scales, demonstrate that the ensemble averages of galaxy properties are reproduced by the simulations despite underlying run-to-run variability. Indeed, while ensemble averages demonstrate minimal variation, the properties of individual galaxies can exhibit significant fluctuations. The following section explores these individual variations in greater detail.

\subsection{Variation in Fiducial simulations without black hole}\label{NOBHFID}

Figure \ref{fig:FID_SFB350_mdm_mstar_50kpc_reg} illustrates the variations in galaxy dark matter mass (left panel) and stellar mass (right panel), respectively, between matched galaxies in the identical simulations. These results pertain to the Fiducial simulation set run without black holes. 

For both, the variations are quantified using the two methods: pooled standard deviation (\textbf{Method 1}, marked with black asterisks) and Mixed Linear Model (\textbf{Method 2}, colored, starting with the label "total"), as described in Section \ref{sec:statistical_analysis}. The total standard deviation (cyan circles) is calculated by summing both the residual error (i.e. variation within each simulation) and the run-to-run variation obtained using Method 2. 
The run-to-run variation is also plotted in the figures (in magenta-x), along with a rough estimate of the shot noise (dashed line). The data evolution over time is plotted with respect to lookback time in gigayears (Gyr).

\begin{figure}
    \centering
    \includegraphics[width=1\linewidth]{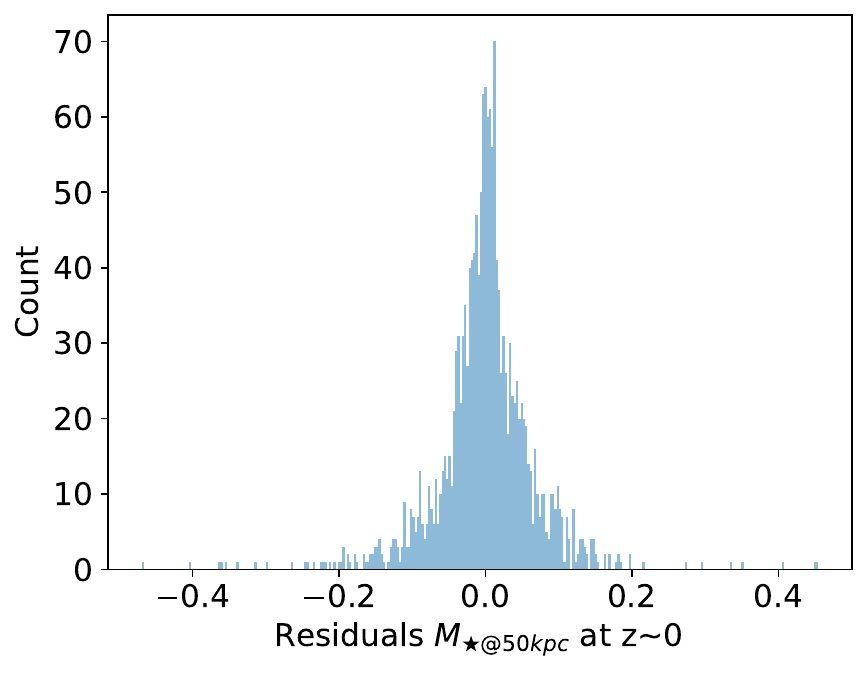}
    \caption{Distribution of residuals in galaxy stellar mass for the Fiducial simulation set at  $z = 0$.}
    \label{fig:residualhisto}
\end{figure}

\begin{figure*}[t]
    \centering
    \begin{minipage}[t]{0.45\textwidth}
        \centering
        \includegraphics[width=\linewidth]{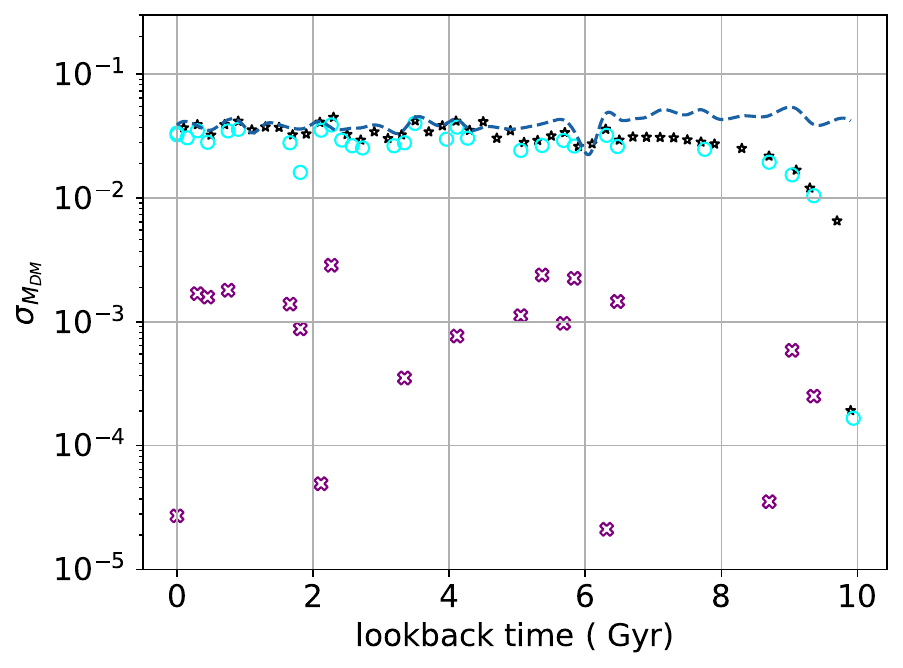}
    \end{minipage}
    \hfill
    \begin{minipage}[t]{0.45\textwidth}
        \centering
        \includegraphics[width=\linewidth]{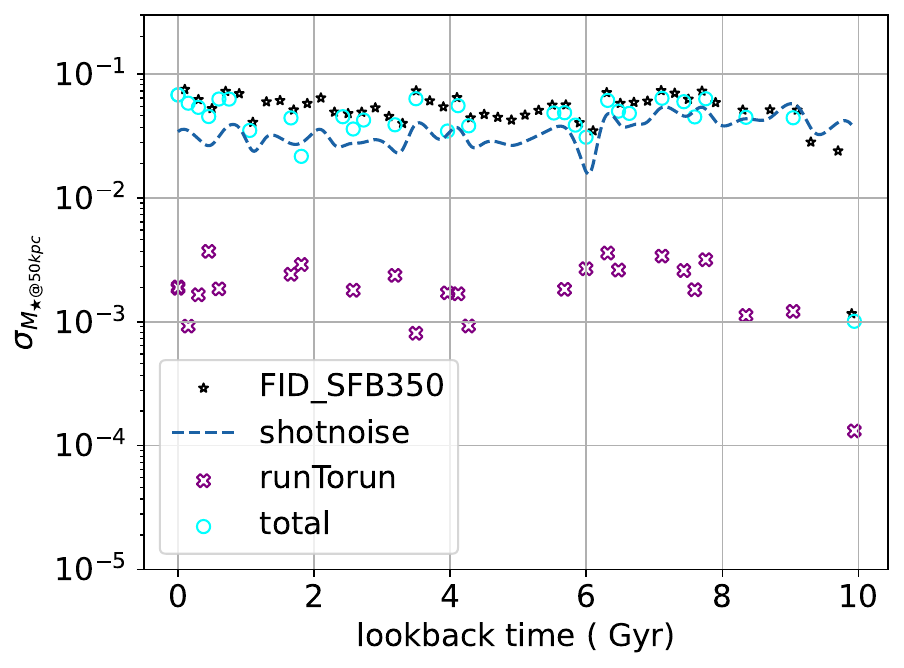}
    \end{minipage}
    \caption{Variation in galaxy dark matter mass (left panel) and stellar mass (right panel: stellar mass is comprised of all stellar particles within an aperture of 50 pkpc from galaxy center) for Fiducial simulations plotted as a function of lookback time in Gyrs. Variation as computed from Method 1 (pooled standard deviation; black stars) and Method 2 (total and run-to-run; cyan-circles and magenta-x respectively) is presented along with an approximate, lower limit estimate of shot noise (dashed colored line). The close overlap between the total variation computed from the two methods is visually apparent and highlights the consistency of the two approaches.}
    \label{fig:FID_SFB350_mdm_mstar_50kpc_reg}
\end{figure*}

Figure \ref{fig:residualhisto} shows that the distribution of residuals for galaxy stellar mass at $z = 0$. The distribution is approximately Gaussian, centered near zero, and displays no strong skewness. This supports the assumption of normally distributed errors required by the Mixed Linear Model and indicates that the model fit is statistically well-behaved.

From Figure \ref{fig:FID_SFB350_mdm_mstar_50kpc_reg}, we observe that the variations computed using the two methods are consistent, although the data exhibit a scattered rather than smooth evolution. Star formation peaks before 10 Gyr; however, for the adopted setup, \textsc{OpenGadget3} with full-physics is not numerically deterministic (although the individual modules for gravity, hydrodynamics, and star formation are). Consequently, minute variations remain negligible until approximately 9 Gyr for the Fiducial model, after which they grow and stabilize at $\sigma_{M_{\text{dm}}}$=0.03-0.05 and $\sigma_{M_*}$=0.04-0.08. In both cases, the total variation closely approaches the level expected from shot noise, indicating that shot noise dominates for these low-resolution runs. This behavior is most evident for $\sigma_{M_{\text{dm}}}$, while $\sigma_{M_*}$ remains slightly above the shot-noise limit.

The higher $\sigma_{M_*}$ reflects the cumulative effects of highly nonlinear and localized processes such as star formation, cooling, gas dynamics, and in particular, feedback mechanisms (as demonstrated in the succeeding section through feedback-related tests). At late times, the variation plateaus, likely driven by gas exhaustion, which suppresses star formation and limits further divergence. This physical constraint explains the observed saturation.
Additionally, we tested the sensitivity of the system to small numerical perturbations at the outset by performing additional runs in which the number of MPI tasks was varied. This varied-MPI test set is conceptually similar to introducing small perturbations through alternative mechanisms (e.g. explicitly introducing small differences in the initial conditions) and provides one approach to probing variability in simulations that may otherwise yield byte-identical results. As expected, the varied-MPI runs show greater divergence, particularly at the peak of star formation, which is not visible in the identical-MPI-task set. However, the system still plateaus similarly to the unperturbed case, due to gas exhaustion regulating the growth of variations (see \ref{sec:appendix_variedMPI}).
In comparison to $\sigma_{M_*}$, the plateauing of dark matter mass aligns closely with the shot noise floor, reflecting the inherently stable nature of collisionless dark matter dynamics.

The run-to-run variation for each of these properties are similar and less than \(\sigma\)=0.005. This is expected that the noise within each set of simulations is very high, considering the simulations are of very low resolution; we do not see any interesting increase in scatter between runs.

\subsection{Feedback tests}
\label{subsec:results_nobhfeedbacktests}

\begin{figure}[htp]
    \centering
    \begin{minipage}[t]{0.45\textwidth}
        \centering
        \includegraphics[width=\textwidth]{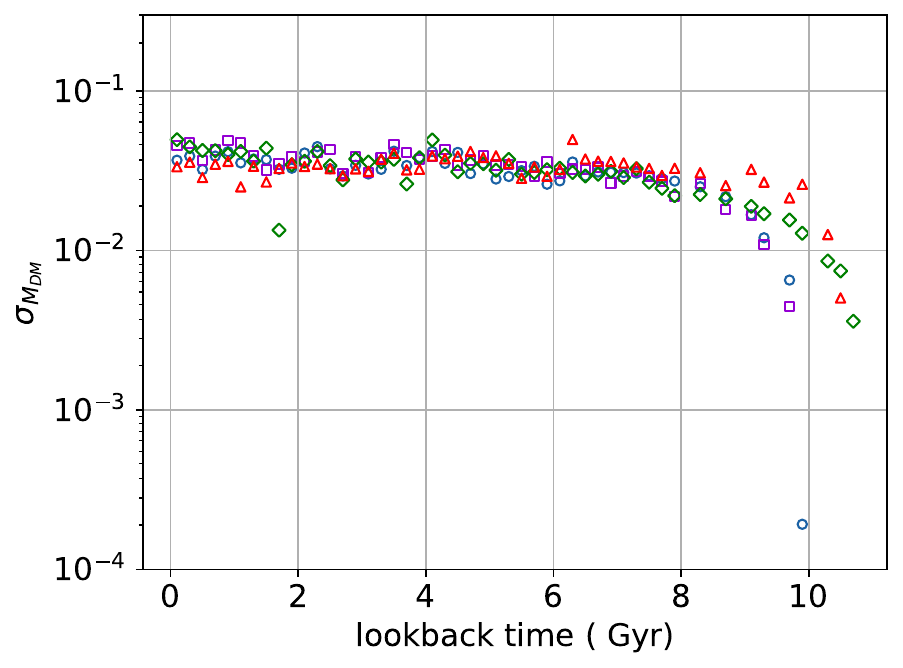}
    \end{minipage}
    \hfill
    \begin{minipage}[t]{0.45\textwidth}
        \centering
        \includegraphics[width=\textwidth]{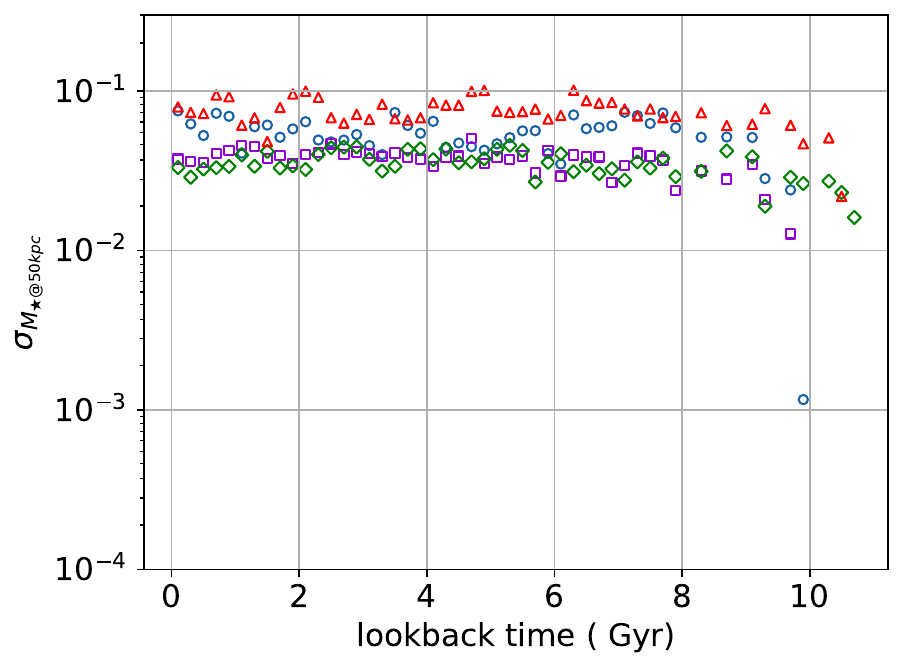}
    \end{minipage}
    \hfill
    \begin{minipage}[t]{0.45\textwidth}
        \centering
        \includegraphics[width=\textwidth]{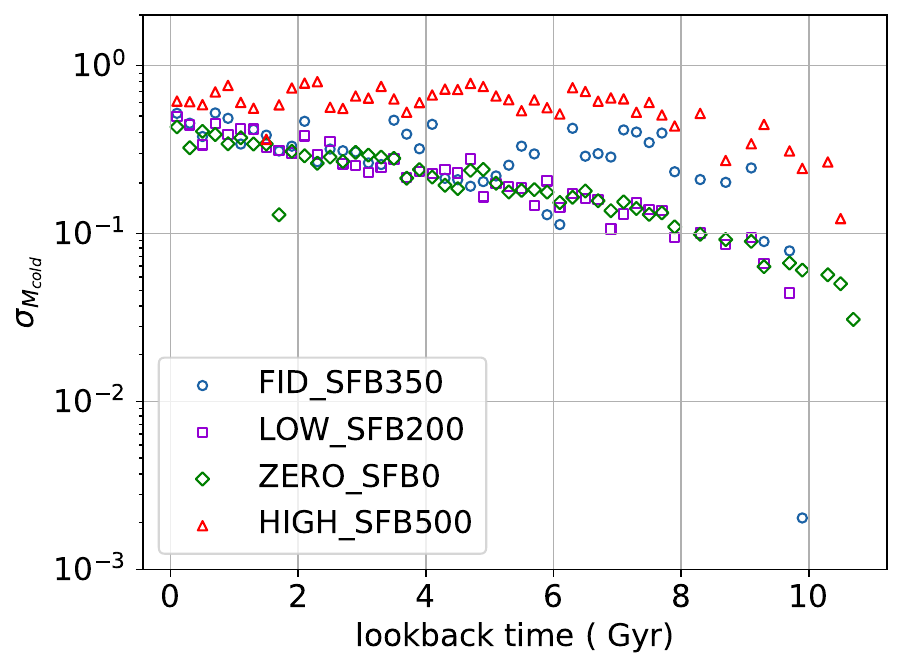}
    \end{minipage}
    \caption{Total variation (Method 1) in galaxy dark matter mass (top panel), stellar mass (middle panel) and cold gas mass (bottom panel) as a function of lookback time (Gyr), shown for the 4 feedback models: Fiducial, LOW, ZERO, HIGH stellar feedback. Each configuration is represented by a distinct color and marker, as indicated in the legend, and this scheme is maintained consistently throughout the paper.}
    \label{fig:FID_SFB350_mdm_mstar_mcold_50kpc_reg}
\end{figure}
Figure \ref{fig:FID_SFB350_mdm_mstar_mcold_50kpc_reg} (top panel) shows the variation in dark matter mass for the 4 flavors of stellar feedback for runs: Fiducial, LOW, ZERO and HIGH as listed in Table \ref{tab:feedback_configs}. The variation is computed with Method 1, i.e. pooled standard deviation. Here we can see that the variation is not very different for each feedback case. The effect of changing stellar feedback on dark matter mass is negligible. Although the scatter in dark matter mass is slightly less pronounced towards later times ($< 2$Gyr) for HIGH feedback than others, all the cases remain more or less the similar, with variation growing from 11-10Gyr and plateauing to a range of $\sigma_{M_{\text{dm}}}$=0.03-0.06 with some scatter.

Figure \ref{fig:FID_SFB350_mdm_mstar_mcold_50kpc_reg} (middle panel) shows the variation in stellar mass for each feedback case, where the total variation is computed with Method 1. The LOW and ZERO runs show reduced scatter compared to the Fiducial (already discussed), whereas HIGH run shows increased scatter, with $\sigma_{M_*}$ between 0.06-0.1.

\paragraph{Why the increase in $\sigma_{M_*}$ for higher stellar feedback?} As already discussed for the Fiducial case, the run-to-run variation is low, \(\sigma_{M_{*}}\) under 0.005. Thus, the increase in $\sigma_{M_*}$ is not due to higher feedback driving an increase in run-to-run variation. Instead, it indicates an increase in noise within each simulation, reflecting higher stochastic variation in each simulation associated with the feedback.

Stellar feedback can lead to "bursty" star formation because feedback from supernovae or winds can temporarily heat and expel gas, quenching star formation in some regions, while other regions may still have gas available to form stars. This leads to increased noise in the stellar mass associated with each simulation. The effect can be further assessed by checking the impact of stellar feedback driving high variation in the cold gas mass as shown in Figure \ref{fig:FID_SFB350_mdm_mstar_mcold_50kpc_reg} (bottom panel). 
The total variation computed with Method 1 shows consistently high variability in the cold gas mass ($\sigma_{M_{\text{cold}}} \approx 0.4-0.8$), with the highest scatter occurring in simulations with stronger stellar feedback, particularly at early times.
This drives subsequent variations in the SFR, which ultimately results in higher variation in galaxy stellar mass in each simulation. Thus high stellar feedback, in simulations without black holes, increases the stochastic variation of a given simulation.

\paragraph{Does that mean that reduced or zero feedback is better in terms of reducing stochastic variation?}
With reduced feedback or lack thereof (cases LOW and ZERO), the only dominant effect is star formation, unregulated, it leads to excessive gas cooling, leading to overproduction of stars and thus rapid depletion of gas. Thus these galaxies show less variability in their star formation histories (albeit unrealistically high SFR), as star formation proceeds smoothly and continuously until all the gas is exhausted. This reasoning is particularly discussed by \cite{keller_chaos_2019} who run dwarf galaxy simulations with and without feedback. They state that for unregulated star formation, variations are dampened, considering the heightened star formation efficiency (SFE); but, the effect does not hold when this one-to-one relation is compromised. They test this by varying the gas fraction for different sets of simulations testing for both cases of with and without stellar feedback. They find that for lower gas fractions, the variation is higher for unregulated star formation and only dampened for certain higher values of gas fraction.  But with stellar feedback, the fluctuations in stellar mass are maintained  regardless of the gas fraction. 

In addition to varying initial gas fractions, the presence of black holes can also disrupt the observed dynamics of high SFE, gas depletion and reduced $\sigma_{M_*}$. The variability introduced by stellar feedback dominates in the absence of additional regulating mechanisms such as black hole feedback, which, when included, can suppress star formation and thereby is expected to influence the overall variability.

\subsubsection{Variation in different galaxy mass bins}
\begin{figure}[htp]
    \centering
    \begin{minipage}[t]{0.45\textwidth}
        \centering
        \includegraphics[width=\textwidth]{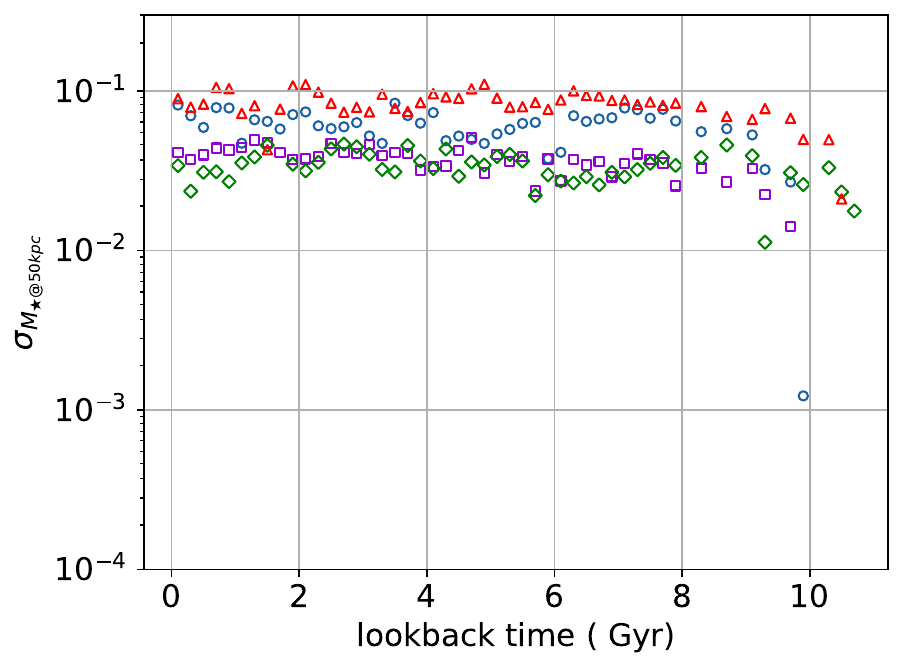}
    \end{minipage}
    \hfill
    \begin{minipage}[t]{0.45\textwidth}
        \centering
        \includegraphics[width=\textwidth]{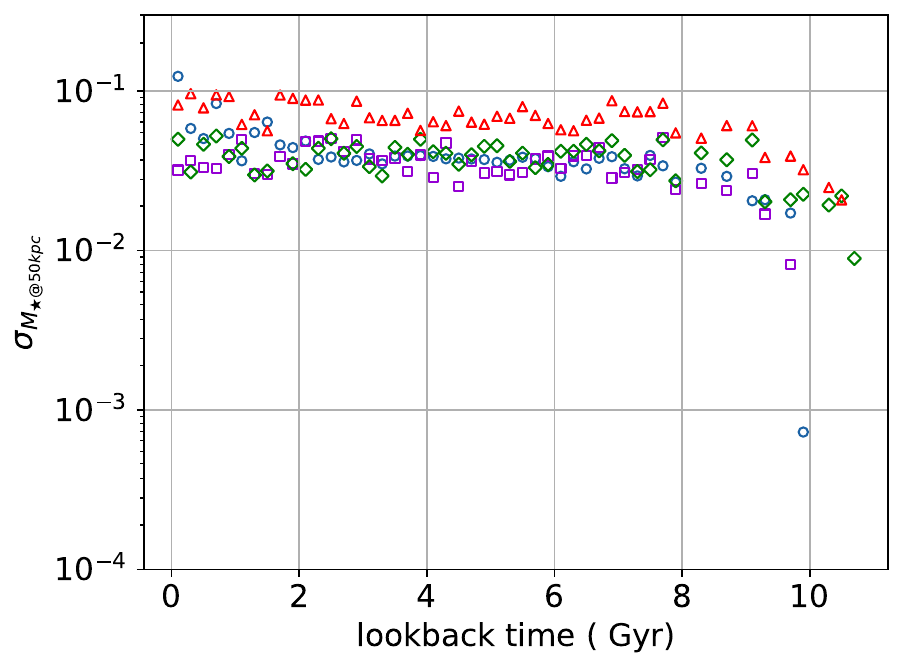}
    \end{minipage}
    \hfill
    \begin{minipage}[t]{0.45\textwidth}
        \centering
        \includegraphics[width=\textwidth]{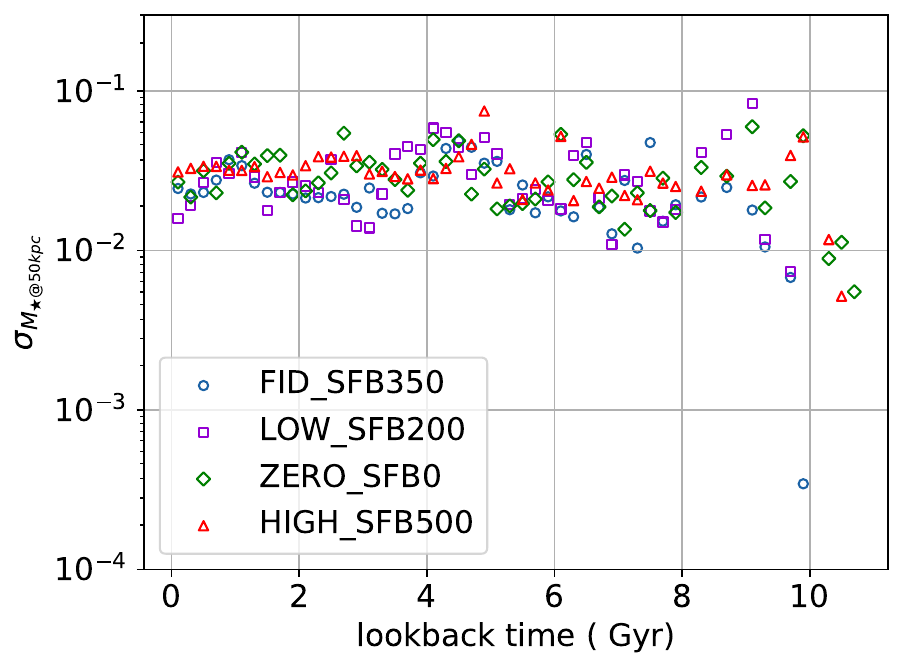}
    \end{minipage}
    \caption{Total variation (Method 1) in galaxy stellar mass as a function of lookback time in Gyrs for galaxies within a mass bin of [\(11.5 < {\rm log} M_{SUB}(M_\odot) < 12.0\)] (top panel), [\(12.0 < {\rm log} M_{SUB}(M_\odot) < 12.5\)] (middle panel) and  [\(12.5 < {\rm log} M_{SUB}(M_\odot) < 13.0\)] (bottom panel) shown for the four feedback models: Fiducial, LOW, ZERO, HIGH stellar feedback. \(M_{SUB}\) denotes the total mass summed from all the particles bound to the galaxy, taken from SUBFIND catalog.}
    \label{fig:D1_1x_NOBH__mstar_50kpc_1_binneddiff}
\end{figure}

The results for \(\sigma_{M_{*}}\) shown so far considers a large distribution of galaxies above \(M_{SUB} > 10^{11} M_\odot\). In this part, we consider how the stellar mass can vary within galaxies belonging to different mass bins. In Figure \ref{fig:D1_1x_NOBH__mstar_50kpc_1_binneddiff}, \(\sigma_{M_{*}}\) is plotted for galaxies belonging to mass bins [\(11.5 < {\rm log} M_{SUB}(M_\odot) < 12.0\)] (top panel), [\(12.0 < {\rm log} M_{SUB}(M_\odot) < 12.5\)] (middle panel) and [\(12.5 < {\rm log} M_{SUB}(M_\odot) < 13.0\)] (bottom panel) respectively. Each time and mass bin always considers a sample size \(> 50\). 

As seen, in Figure \ref{fig:D1_1x_NOBH__mstar_50kpc_1_binneddiff} top and middle panel, the results of the full distribution (Figure \ref{fig:FID_SFB350_mdm_mstar_mcold_50kpc_reg} (middle panel)) is still reflected. Thus for the galaxies of relatively lower mass bins [\({\rm log} M_{SUB}(M_\odot) < 12.5\)], the stochastic variation is higher for stronger feedback compared to LOW and ZERO feedback.
But, when it comes to galaxies belonging to higher end of the mass bins [\({\rm log} M_{SUB}(M_\odot) > 12.5\)], see bottom panel of Figure \ref{fig:D1_1x_NOBH__mstar_50kpc_1_binneddiff} (containing galaxies of mass bins: [\(12.5 < {\rm log} M_{SUB}(M_\odot) < 13.0\)]), the variation for all the feedback types are scattered but with $\sigma_{M_*}$ more or less similar without a significant trend. 
This result reflects the differential impact stellar feedback has on galaxies depending on their mass. Lower-mass galaxies experience a stronger impact from stellar feedback due to shallower gravitational potential wells; the energy and momentum of supernovae feedback or winds can drive gas out of the galaxy more efficiently, thus limiting the available gas for star formation and therefore having an impact on the galaxy's stellar mass. The varying trends in $\sigma_{M_*}$ with respect to the feedback is a reflection of this.

On the contrary, higher-mass galaxies with deeper gravitational potential wells are less affected by stellar feedback. Although stellar feedback still plays a role, it is considerably less efficient in driving out significant amounts of gas. Thus, for the relatively larger mass bin [\(12.5 < {\rm log} M_{SUB}(M_\odot) < 13.0\)], the effect of HIGH feedback is not evident and thus the stochastic variation is more or less similar to a lower or unregulated star formation.

\section{Variability including black holes}\label{sec:results_withbh}

\begin{figure*}[t]
    \centering
    \begin{minipage}[t]{0.45\textwidth}
        \centering
        \includegraphics[width=\linewidth]{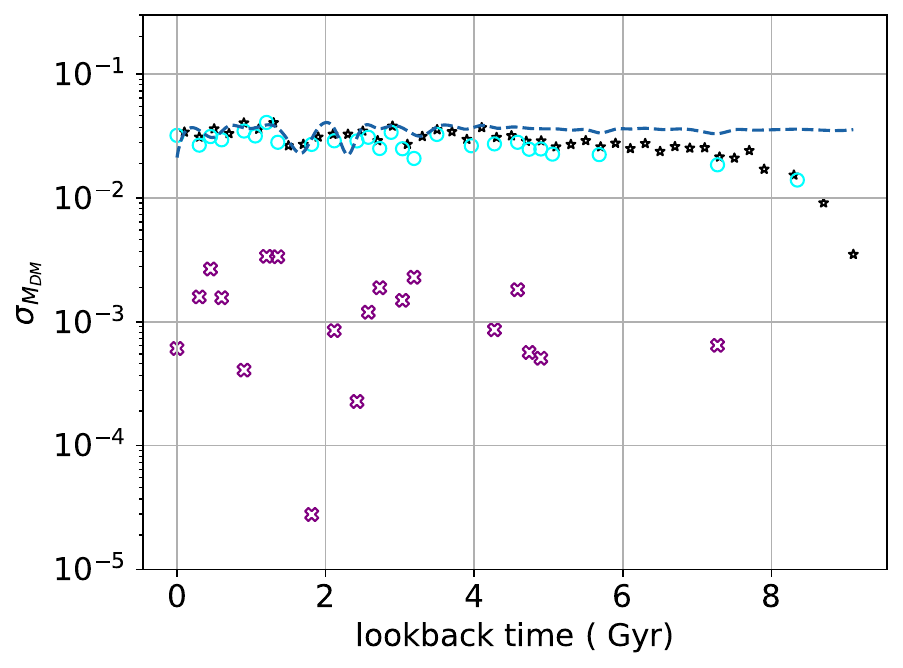}
    \end{minipage}
    \hfill
    \begin{minipage}[t]{0.45\textwidth}
        \centering
        \includegraphics[width=\linewidth]{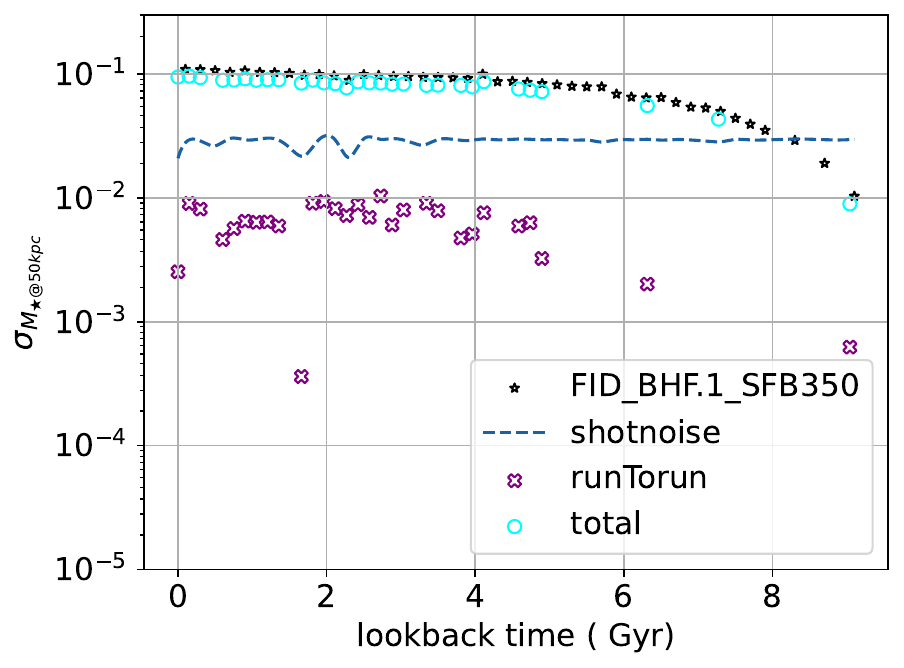}
    \end{minipage}
    \caption{Same as Figure \ref{fig:FID_SFB350_mdm_mstar_50kpc_reg}, but for Fiducial simulations including black holes. Variation in galaxy dark matter mass (left panel) and stellar mass (right panel) is shown.}
    \label{fig:FID_BHF.1_SFB350_mdm__mstar_50kpc__reg}
\end{figure*}

Figure \ref{fig:FID_BHF.1_SFB350_mdm__mstar_50kpc__reg} shows results analogous to those in Section \ref{NOBHFID} (i.e. Figure \ref{fig:FID_SFB350_mdm_mstar_50kpc_reg}), but for simulations including black holes.
Here, total variation (Method 2: sum of residual and run-to-run variation, in cyan-circles), run-to-run variation (in magenta-x) is represented for the Fiducial simulations with black holes for galaxy dark matter (left panel) and stellar mass (right panel). Also, the variation computed with Method 1 (black stars) is provided along with an approximate lower limit estimate of shot noise (colored dashed line).

Variations increase steadily from \(\approx\) 9 Gyr, plateauing at $\sigma_{M_{\text{dm}}}$ \(\approx\) 0.04 and $\sigma_{M_*}$ \(\approx\) 0.1 thereafter (corresponding to \(\approx 10 \%\) and \(\approx 25 \%\) scatter, respectively). The total variation computed using both methods is consistent. For dark matter mass, the total variation closely approaches the lower limit set by shot noise, similar to the no-black-hole (NOBH) case.

For stellar mass, shot noise does not appear to be the sole contributor to the total variation at later times (considering the total stellar mass variation is systematically higher than in the corresponding NOBH case from \(\approx\) 6 Gyr onward). The run-to-run component for both properties remains negligible compared to the total variation, with \(\sigma \lesssim 0.01\).

The increased variations are driven by the inclusion of black hole subgrid physics, which is, no doubt, a significant source of variation. This is considering that black hole positions are highly sensitive to the gravitational interaction with surrounding particles, which can even lead to phenomena such as the wandering black hole~\citep{ragone-figueroa_bcg_2018, damiano_dynamical_2024}. This can introduce a significant source of variation that can affect variations in stellar mass buildup. Additionally, both star formation and black hole accretion compete for gas; consequently, we can expect an increased sensitivity of stellar mass to local variation, and thus we observe higher scatter compared to simulations without black holes.

Note that across all properties examined here (including the baseline variation for our NOBH-tests), the residual variance dominates the run-to-run variance by at least an order of magnitude. Thus, considering that our low-resolution runs are noise-dominated: repeated runs converge to a well-defined expectation value for each galaxy property.
By contrast, if the run-to-run component were to become non-negligible, the interpretation would change qualitatively. In that regime, each individual trajectory realized by the simulation should be regarded as equally valid.
This distinction between noise-dominated and run-to-run variance has direct consequences for whether individual galaxies can be robustly studied within a given numerical setup.
In the present work, we do not attempt to define a quantitative threshold for this transition (for e.g., in terms of the relative effect size or the fraction of run-to-run variance relative to the total variance), since this requires systematic resolution-scaling tests, which we defer to future work. Notably, the run-to-run component derived from Method 2 represents a statistical variance rather than a purely physical divergence; it can itself respond to changes in the underlying noise floor. Its relative effect size therefore provides a useful indicator of when repeated-run variability becomes non-negligible. Nonetheless, it is important to emphasize that determining whether variability is noise dominated or run-to-run dominated is scientifically significant: in the former case ensemble averaging converges to a true expectation value, while in the latter case ensemble statistics must be interpreted probabilistically, reflecting fundamental limits on predictability at the level of individual galaxies or sub-samples.

\subsection{Feedback tests}
\begin{figure}
    \centering
    \includegraphics[width=1\linewidth]{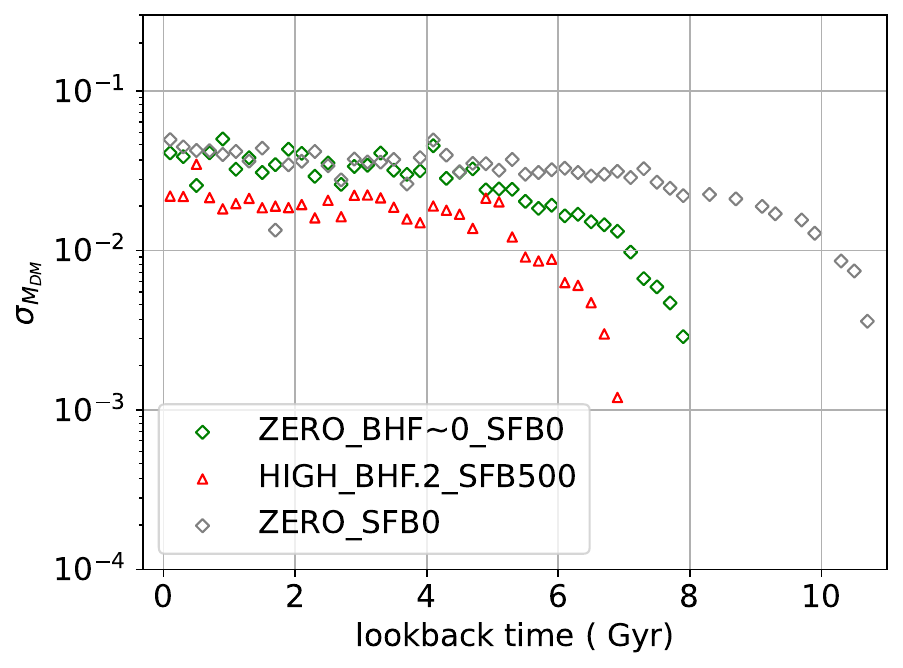}
    \caption{Total variation (only Method 1) in galaxy dark matter mass as a function of lookback time (Gyr), shown for the feedback models: ZERO, HIGH feedback with black holes (colored lines and markers as indicated in the legend), and ZERO feedback without black hole (in gray) provided for comparison.}
    \label{fig:D1_1x_NOBH__fbmdm_pairDiff}
\end{figure}

Figure \ref{fig:D1_1x_NOBH__fbmdm_pairDiff} shows the total variation in galaxy dark matter mass (computed with Method 1) from the feedback tests including black holes for ZERO, HIGH feedback tests. ZERO feedback case without black hole is provided for reference (see Table \ref{tab:feedback_configs} for the respective feedback parameter values used).

The variation grows from around 7-8 Gyrs and plateaus to \(\sigma_{M_{dm}} \approx 0.02-0.06\). The plot clearly show that with stronger feedback, there is a reduction in the scatter in the galaxy dark matter mass;  $\sigma_{M_{\text{dm}}}$ in HIGH plateaus below 0.03 with some scatter. Both the cases of zero feedback with and without black hole have similar trends with  $\sigma_{M_{\text{dm}}}$ above 0.03 up to 0.06 with some scatter.

\begin{figure}
    \centering
    \includegraphics[width=1\linewidth]{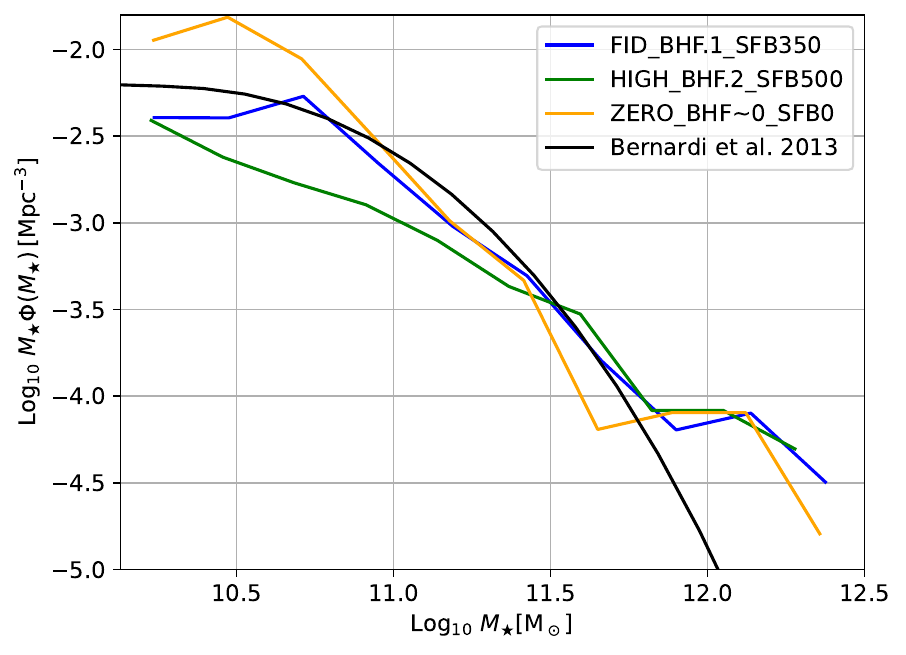}
    \caption{GSMF for the feedback models: FID, ZERO, HIGH feedback at  $z = 0$. GSMF is normalised as described in \cite{bassini_dianoga_2020}. Observational data from Bernardi et al. (2013) in black.}
    \label{fig:r3_stellarMassFunction}
\end{figure}

For the specific choice of feedback parameters in the HIGH feedback case, the impact on the galaxy population is most pronounced at the low-mass end ($M \lesssim 10^{12} \, M_{\odot}$). As shown in Figure \ref{fig:r3_stellarMassFunction}, where the GSMF is plotted for the FID, HIGH, and ZERO runs along with the \cite{bernardi_massive_2013} observational data provided for reference; considering the stellar mass ascribed by SUBFIND to all the galaxies in a single run in each respective test. GSMF is normalized following \cite{bassini_dianoga_2020}. As can be observed in the Figure, the HIGH feedback case exhibits the characteristic flattening of the low-mass slope (\cite{dekel_origin_1986, larson_effects_1974, white_galaxy_1991}) much more prominent than Fiducial.

\begin{figure}
    \centering
    \includegraphics[width=1\linewidth]{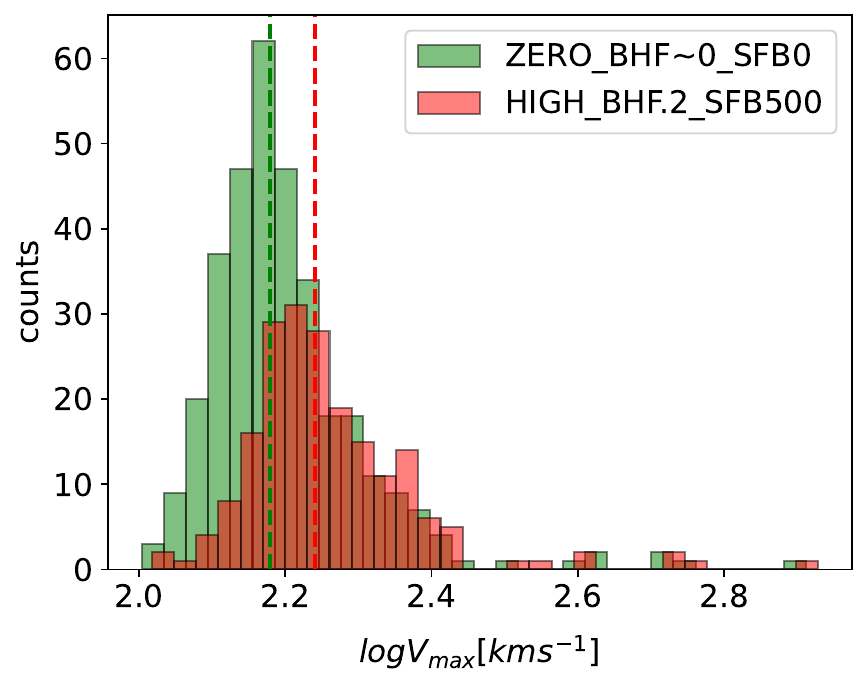}
    \caption{Histogram of maximum circular velocities at  $z = 0$  for BCGs in one of the runs of the ZERO and HIGH feedback respectively. Dashed vertical lines indicate the mean \(V_{\max}\) for each distribution.}
    \label{fig:D1_1x_vmax_histogram}
\end{figure}

This selection effect results in the sample of HIGH-galaxies being, on average, denser than in the other cases (see Figure \ref{fig:D1_1x_vmax_histogram} for a comparison of maximum circular velocity between HIGH and ZERO). The maximum circular velocity is defined as \(V_{\max} = \max \left[ \sqrt{GM(<r)/r} \right]\) as function of radius r; \(M(<r)\) is the enclosed subhalo mass and G is the gravitational constant. This bias effectively lowers the shot noise in the sample and thus reduces $\sigma_{M_{\text{dm}}}$.

\begin{figure*}[t]
    \centering
    \begin{minipage}[t]{0.45\textwidth}
        \centering
        \includegraphics[width=\linewidth]{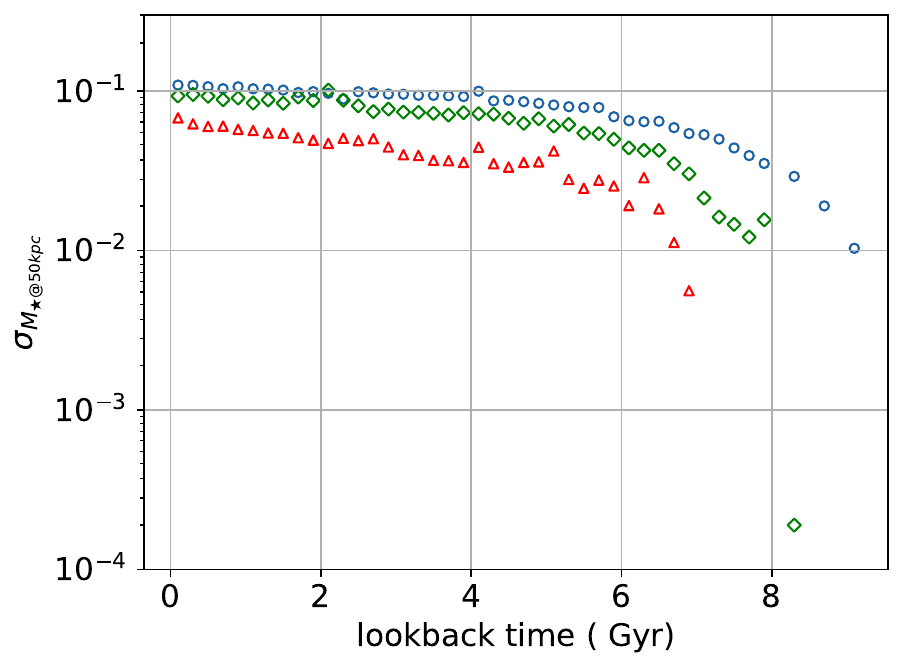}
    \end{minipage}
    \hfill
    \begin{minipage}[t]{0.45\textwidth}
        \centering
        \includegraphics[width=\linewidth]{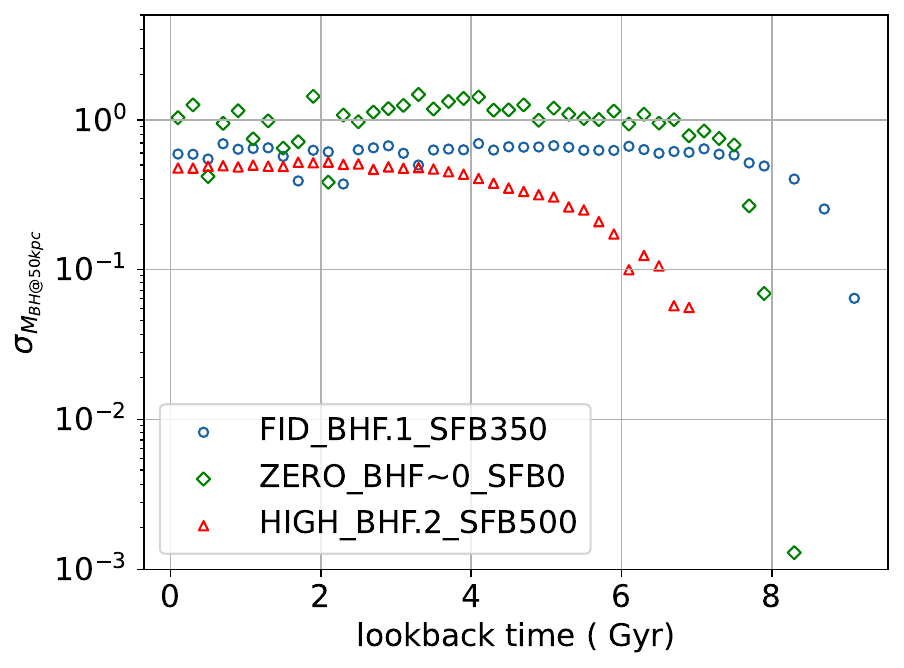}
    \end{minipage}
    \caption{Total variation (Method 1) in galaxy stellar mass (left panel) and black hole mass (right panel) as a function of lookback time (Gyr), shown for the feedback models: FID, ZERO and HIGH.}
    \label{fig:fbtest_D1_1x__mstar_50kpc_AndD1_1x__snapmbh_}
\end{figure*}

For stellar mass results, Figure \ref{fig:fbtest_D1_1x__mstar_50kpc_AndD1_1x__snapmbh_} (left panel) shows the total variation in galaxy stellar mass computed using Method 1 for the FID, ZERO, and HIGH feedback test cases. The ZERO feedback case exhibits variability comparable to the FID run. In contrast, the HIGH feedback case shows reduced variation at earlier epochs ($\ge 5$ Gyr), remaining below the FID and ZERO cases throughout. While the FID and ZERO runs approach a plateau of $\sigma_{M_*}$= 0.1 at $z = 0$, the HIGH feedback case continues to increase.

The scatter in \( \sigma_{M_*} \) is influenced by both stellar and black hole feedback.
Results from the no-black-hole-feedback tests (Section \ref{subsec:results_nobhfeedbacktests}) show that unregulated star formation leads to rapid gas exhaustion, resulting in a reduced \( \sigma_{M_*} \). However, when black holes are included, this trend is reversed: the high-feedback case exhibits lower \( \sigma_{M_*} \) than the Fiducial setup. More efficient feedback better suppresses the scatter in star formation. Each feedback scenario eventually reaches a plateau in \( \sigma_{M_*} \), which coincides with the onset of gas exhaustion. The timing and extent of this exhaustion are jointly controlled by stellar and black hole feedback. 
When gas is exhausted, star formation is quenched, leading to a convergence and stabilization of \( \sigma_{M_*} \) across the different runs (Gas is not as exhausted for HIGH feedback case as is the case for other cases, thus $\sigma_{M_*}$ does not plateau).

An inverse correlation can be observed between the black hole mass scatter, $\sigma_{M_{\text{bh}}}$, and the strength of black hole feedback in Figure \ref{fig:fbtest_D1_1x__mstar_50kpc_AndD1_1x__snapmbh_} (right panel). In simulations with more efficient feedback, the final black hole masses exhibit significantly reduced scatter. For instance, in the HIGH-feedback configuration, $\sigma_{M_{\text{bh}}}$ is constrained to \(\approx 0.5\), whereas in the unregulated feedback scenario, \( \sigma_{M_{\text{bh}}} \approx 1 \). Although the latter represents an unphysical extreme, it highlights the critical role of feedback in regulating black hole growth. Despite the scatter in instantaneous accretion rates being comparable across different feedback strengths, the cumulative effect of suppressed feedback leads to excessively high accretion over cosmic time (See Figure \ref{fig:fbtest_D1_1x__mcold_50kpc_AndD1_1x__bhmd_} right panel in \ref{sec:appendix_withbh_sfrbhmd} as well as variation in galaxy cold gas mass provided in the left panel). Consequently, black holes in the ZERO-feedback case grow to masses that are, on average, two orders of magnitude larger than those in regulated runs. The unrealistic massive ZERO-black holes accrete all of the cold gas, thereby suppressing the star formation. This results in a reduction of $\sigma_{M_*}$ which explains why $\sigma_{M_*}$ in the ZERO-feedback case is not as elevated as might otherwise be expected (Figure \ref{fig:fbtest_D1_1x__mstar_50kpc_AndD1_1x__snapmbh_} left panel).

Further details: the overall trend in the ZERO-feedback case reflects two competing effects. For galaxies with halo masses in the range [\(12.0 < {\rm log} M_{SUB}(M_\odot) < 13.0\)], black holes grow sufficiently massive to strongly suppress star formation, lowering the apparent scatter. In this regime, the behavior resembles the ZERO-NOBH or LOW-NOBH case, though here the suppression arises from the high black hole accretion and gas exhaustion rather than from unregulated star formation. At lower halo masses [\( M_{SUB}(M_\odot) < 12.0\)], star formation still competes effectively for the available gas. This manifests as a set of outliers in the ZERO-feedback runs that drive $\sigma_{M_{\text{bh}}}$ to higher values, even though the bulk of the distribution remains comparable to the HIGH and FID cases (with \( \sigma_{M_{\text{bh}}} \approx 0.5\) ).

\subsection{Change in precision}\label{mixedprecision}
The choice of numerical precision in simulations reflects a balance between computational cost and the control of accumulated round-off error. In systems where higher-order terms or conservation properties play an important role, round-off error becomes increasingly relevant as the number of operations increases.
For instance, in idealised tests of SPH density estimation, the intrinsic spatial accuracy of the method (e.g. \(\approx 0.1\%\) density reconstruction accuracy with sufficient neighbours; see \cite{dehnen_improving_2012}) might suggest that lower numerical precision is adequate for some operations. However, this conclusion does not account for long-term integration effects in a complex, coupled system.

Our work utilizes cosmological simulations that are non-linear and feature a significantly more complex numerical setup including coupled gravity, adaptive time-stepping, a modern SPH solver, and stochastic subgrid models. In this environment, small numerical differences arising from round-off error compound over thousands of integration steps. To ensure numerical stability and minimize accumulated integration error, higher precision arithmetic is therefore required for key quantities.

Throughout the preceding sections of this study, we have utilized the full double precision (64-bit) configuration as our Fiducial baseline. Our rationale for this choice is to establish a more numerically stable environment. Having established this baseline, we now investigate the impact of a more computationally efficient mixed precision setup.

Mixed precision (MP) refers to the use of multiple floating point precision levels, typically, single precision (32-bit) and double precision (DP) within the same computational workflow. The purpose of this approach is to balance numerical accuracy with computational efficiency, given that some operations are performed in lower precision to significantly reduce memory usage and to increase speed. Critical calculations requiring higher numerical stability retain double precision: e.g. particle positions, variables in SPH calculations such as entropy, pressure gradients. Gravitational force calculations (accumulation of forces in long range and direct summation), other quantities like particle mass, velocity updates are in single precision.

Given its advantages in computational speed and reduced memory footprint, mixed precision arithmetic is currently widely adopted in simulations using \textsc{OpenGadget3}. In this section, we investigate how mixed precision impacts variability in simulation outcomes.

\begin{figure*}[t]
    \centering
    \begin{minipage}[t]{0.45\textwidth}
        \centering
        \includegraphics[width=\linewidth]{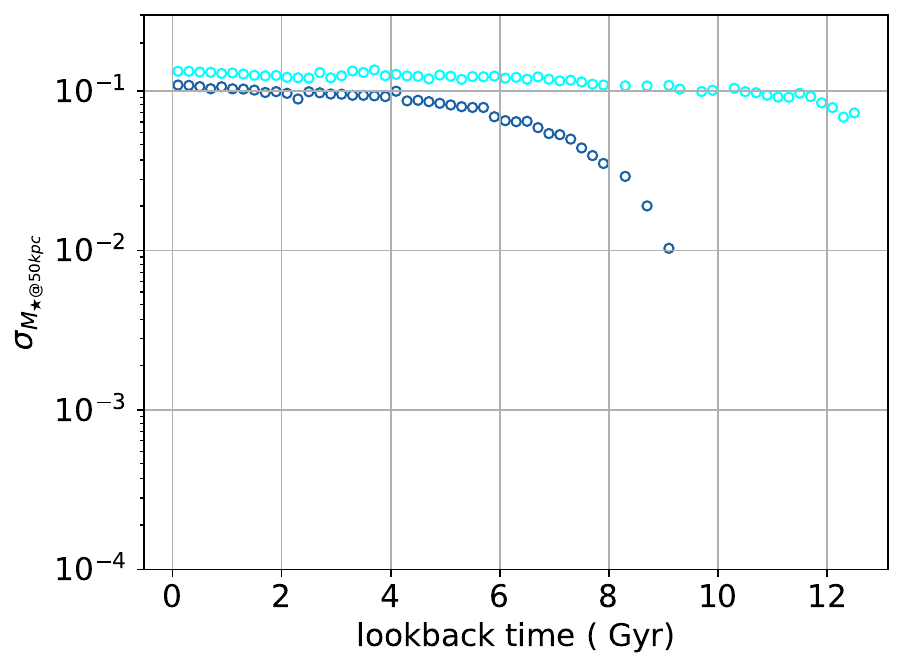}
    \end{minipage}
    \hfill
    \begin{minipage}[t]{0.45\textwidth}
        \centering
        \includegraphics[width=\linewidth]{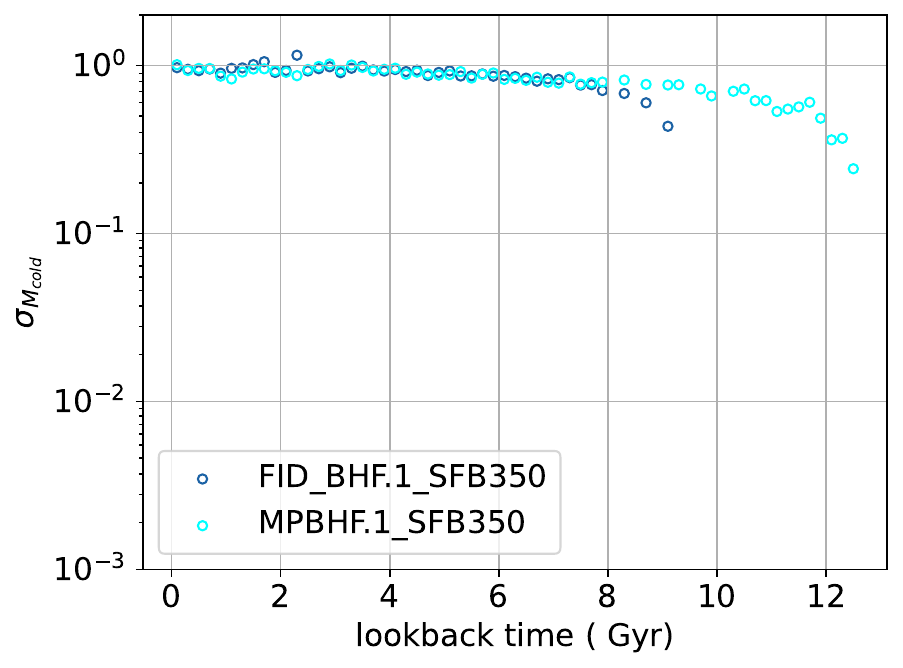}
    \end{minipage}
    \caption{Total variation (Method 1) in galaxy stellar mass (left panel) and cold gas mass (right panel); as a function of lookback time (Gyr), shown for Fiducial double (in blue) and mixed precision sets (MP, in cyan).}
    \label{fig:D1_1x_MP_star_mcold}
\end{figure*}

Figure \ref{fig:D1_1x_MP_star_mcold} compares variation in galaxy stellar mass and cold gas mass (variation in galaxy dark matter mass and black hole mass provided in Figure \ref{fig:D1_1x__mdm_MP_MBH}) for the Fiducial simulation including black holes run in double precision (already discussed) with its mixed precision counterpart. In addition to the change in the numerical precision, the snapshots in the latter are also stored with reduced numerical precision.

The difference in round-off errors between the two simulations sets are evident with the errors accumulating more rapidly in the mixed precision runs well before $\approx 9\,\mathrm{Gyr}$. 
$\sigma_{M_*}$ (Figure \ref{fig:D1_1x_MP_star_mcold} left panel) plateaus to 0.1; similar to, but with a slight enhancement over the double precision scatter. Note that we do not observe a corresponding increase in the variation of cold gas mass (Figure \ref{fig:D1_1x_MP_star_mcold} right panel, $\sigma_{M_{\text{cold}}} \approx 1$) or in the instantaneous star formation rates.

\begin{figure*}[t]
    \centering
    \begin{minipage}[t]{0.45\textwidth}
        \centering
        \includegraphics[width=\linewidth]{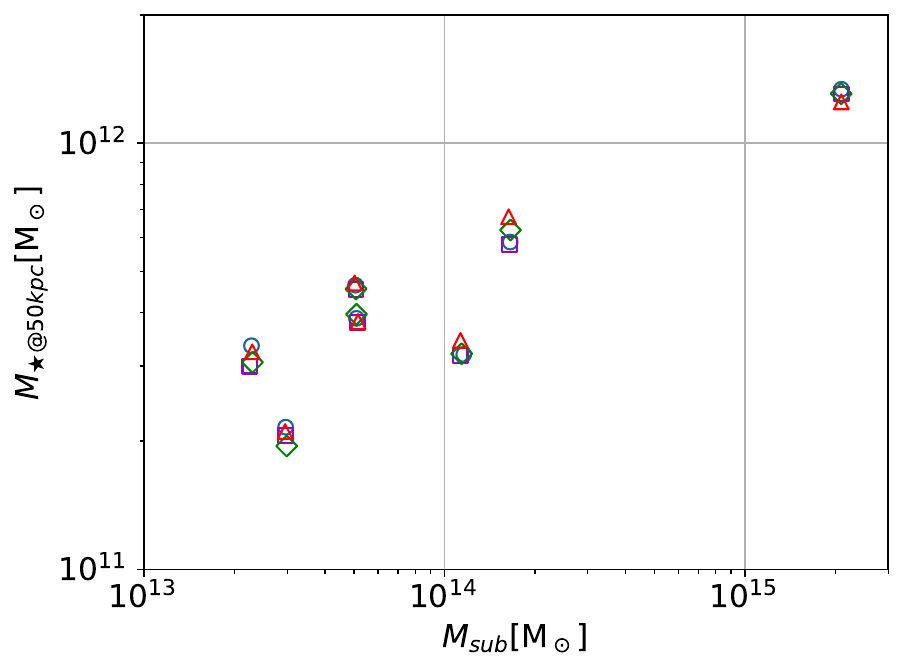}
    \end{minipage}
    \hfill
    \begin{minipage}[t]{0.45\textwidth}
        \centering
        \includegraphics[width=\linewidth]{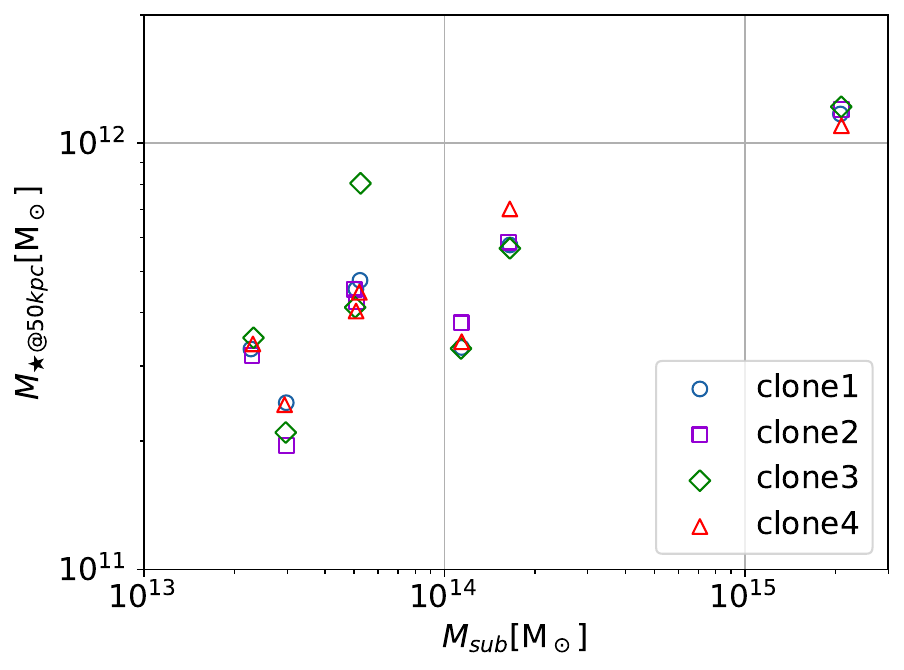}
    \end{minipage}
    \caption{Stellar mass versus subhalo mass correlation for few of the most massive BCGs in Fiducial double precision runs (left panel) and mixed precision runs  (right panel) at $z=0$. Each identical run in the respective set is represented by a different marker and color.}
    \label{fig:D1_1x_FID_BHF.1_SFB350_AND_MPBHF.1_SFB350mstar_50kpc_92var}
\end{figure*}

\begin{figure*}
    \centering
    \includegraphics[width=0.95\textwidth]{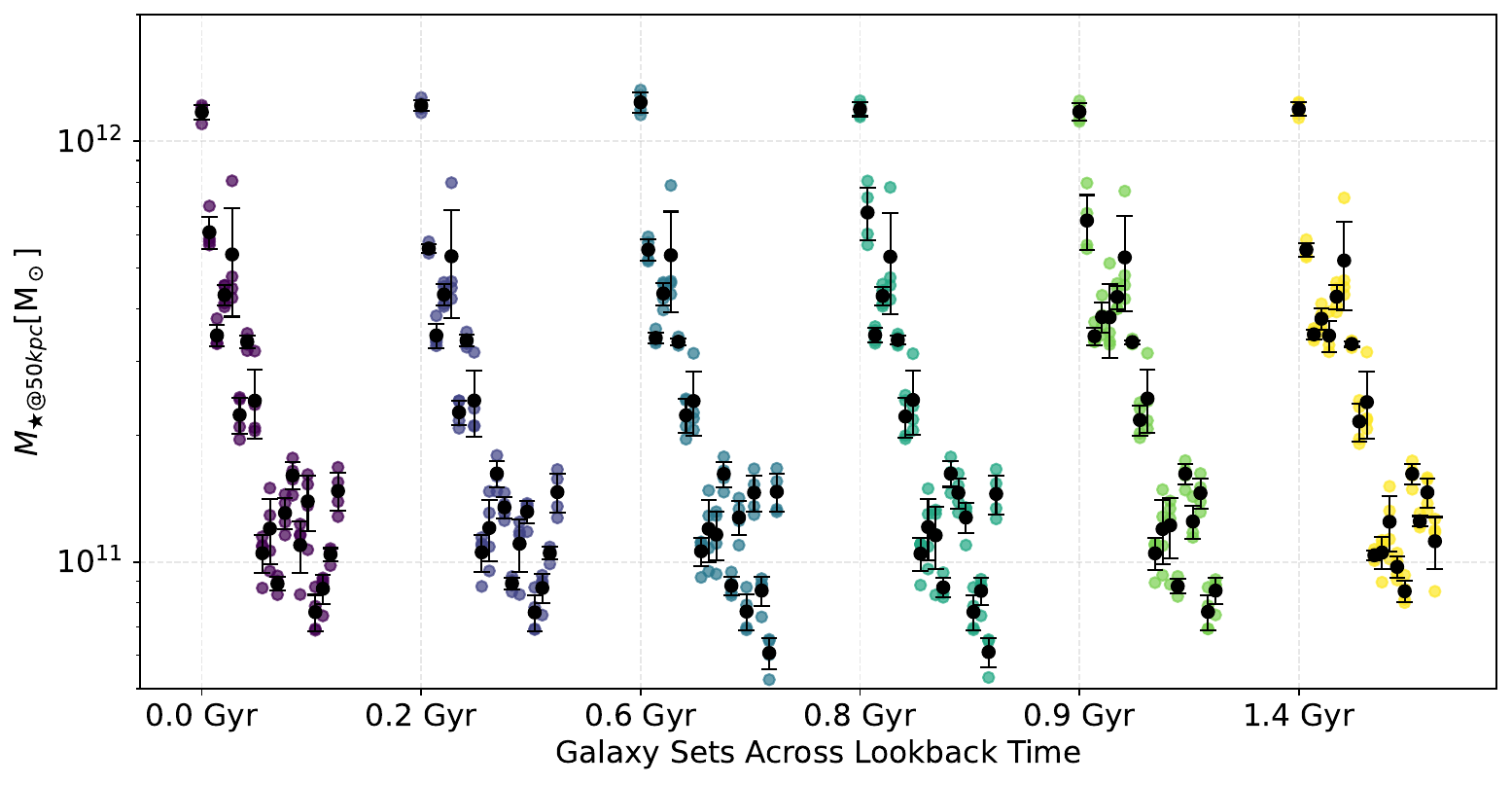}
    \caption{Stellar masses of the N most massive galaxies across a $\sim$1.4 Gyr interval, sampled at five snapshots for mixed precision simulations. Matched galaxies are plotted and for each matched galaxy set, the points indicate the stellar mass from the representative clones, while the vertical error bars roughly illustrate the variability between each set. Different colors correspond to individual galaxies, and the same color is maintained across snapshots to allow visual tracking of whether a galaxy's relative scatter persists over time or fluctuates. This highlights persistent differences (rather than fluctuating) in stellar mass in matched galaxies across identical simulations.
    }
    \label{fig:D1_1x_MPBHF.1_SFB350_nvarsgyr}
\end{figure*}

An enhancement in mixed precision $\sigma_{M_*}$ can be easily visualized in 
Figures \ref{fig:D1_1x_FID_BHF.1_SFB350_AND_MPBHF.1_SFB350mstar_50kpc_92var}, 
where the stellar masses of the most massive BCGs at $z=0$ are shown for each realization in both the DP (Figure \ref{fig:D1_1x_FID_BHF.1_SFB350_AND_MPBHF.1_SFB350mstar_50kpc_92var} left panel) and MP (Figure \ref{fig:D1_1x_FID_BHF.1_SFB350_AND_MPBHF.1_SFB350mstar_50kpc_92var} right panel) test sets.
Compared to DP, the increased scatter among the fourth most massive galaxies is starkly evident in case of MP runs. The time evolution of this divergence is illustrated in Figure \ref{fig:D1_1x_MPBHF.1_SFB350_nvarsgyr}. Here, a selection of the most massive galaxy stellar masses are plotted,  each presented with vertical error bars indicating the variation among clones over a $\sim$1.4 Gyr interval in the MP runs (different colors maintained for galaxies across the snapshots). The figure demonstrates that, in this case, the divergence in stellar mass is largely persistent over time, suggesting a systematic growth of variability.

The effect occurs because one clone (Clone 1) hosts a black hole at \(z\approx 3.42\), while another (Clone 2) does not. When Clone 2 is finally seeded (at \(z\approx 2.31\)), its black hole accretes rapidly and largely catches up to the other clones by the end of the simulation, with final black hole masses remaining of the same order.
In contrast, the impact on stellar mass is more pronounced due to the extended period during which Clone 2 evolves without a black hole. Stellar mass variation increases during this interval and continues until the black hole in Clone 2 grows sufficiently massive (\(z\approx 1.7\)) for feedback to stymie further growth. By this stage, the stellar mass in Clone 2 has already grown substantially relative to the other clones. As a result, \(\sigma_{M_{bh}}\) remains relatively unaffected compared to the corresponding higher variation in stellar mass (see Figure \ref{fig:D1_1x__mdm_MP_MBH}, left panel).

Though outwardly an interesting case, the root cause of this divergence has been attributed to a too-aggressive precision truncation in the Barnes-Hut MPI exchange of accelerations. The newer version of \textsc{OpenGadget3} alleviates this issue by adopting a larger number of double precision quantities compared to our older version of the code. We specifically carried out tests in mixed precision, since this mode is widely used and forms the basis of the long-standing "PROD" version of the code employed in many previous studies. The significantly large divergence observed in this configuration implies that results from such studies are also affected; thus, while large-scale properties remain robust, 'error bars' for individual object properties would be broader compared to double precision simulations. Consequently, researchers analyzing such simulations of individual galaxies must exercise caution, as their results are subject to greater numerical sensitivity and potentially larger uncertainties than those found in higher precision setups.
The recently implemented change is expected to mitigate this behavior, although we have not independently verified whether they are directly related to the divergence we observe.

\section{Summary and conclusions}\label{sec:summaryConclusions}

We investigated variability in galaxy cluster simulations using \textsc{OpenGadget3} code. Identical simulations at the lowest resolution setting (1x) for one of the Dianoga galaxy clusters were executed. Individual galaxies (only central galaxies) were matched across these runs, and variations in their properties were analyzed using two statistical methods: pooled standard deviation (Method 1) and a mixed linear model (Method 2).  Here we list some of our key results:

\paragraph{Without black holes}

\begin{itemize}
    \item For our \textit{Fiducial} model: We find variations of ($\sigma_{M_{\text{dm}}} = 0.03\text{–}0.05$) and ($\sigma_{M_*} = 0.04\text{–}0.08$) (Figure \ref{fig:FID_SFB350_mdm_mstar_50kpc_reg}; relatively higher variation in stellar mass is attributed to shot noise and the influence of highly non-linear processes like star formation and feedback). 
    \item Feedback tests (see Table \ref{tab:feedback_configs}): Negligible effect on the variation in dark matter mass (Top panel of Figure \ref{fig:FID_SFB350_mdm_mstar_mcold_50kpc_reg}) but significantly impacts the variation in stellar mass (Middle panel of Figure \ref{fig:FID_SFB350_mdm_mstar_mcold_50kpc_reg}). Higher stellar feedback has increased scatter ($\sigma_{M_*} \approx 0.08\text{–}0.1$) due to increased stochastic variation (results consistent with \cite{keller_chaos_2019}). 
\end{itemize}

\paragraph{With black holes}
\begin{itemize}
    \item For our \textit{Fiducial} model: yields a $\sigma_{M_{\text{dm}}} \approx 0.04$ (Figure \ref{fig:FID_BHF.1_SFB350_mdm__mstar_50kpc__reg}; consistent with its Fiducial no-black-hole counterpart). However, $\sigma_{M_*}$, saturates at a significantly higher value of \(\approx 0.1\), indicating increased sensitivity of star formation in the presence of black holes.
    \item Feedback tests (see Table \ref{tab:feedback_configs}):
    The HIGH-feedback model exhibits a reduced scatter, with \(\sigma_{M_*} \approx 0.07-0.08\) and a slight decrease in \(\sigma_{M_{\text{bh}}} \approx 0.5\), compared to the Fiducial model ($\sigma_{M_*}$= 0.1 and \(\sigma_{M_{\text{bh}}} \approx 0.6\); see Figure \ref{fig:fbtest_D1_1x__mstar_50kpc_AndD1_1x__snapmbh_}).
\end{itemize}

Our results indicate the regulatory role of feedback in moderating both star formation and black hole growth. However, disentangling the relative contributions of feedback, subgrid physics, and stochastic effects to the overall variability is inherently complex. For instance, \cite{genel_quantification_2019} reported enhanced variability in simulations with feedback, though part of this trend may reflect differences in the inclusion of black holes between their feedback and no-feedback runs. Further parameter exploration will be valuable to better characterize these dependencies and test the generality of our findings.

Run-to-run variation is minimal ($\sigma \lesssim10^{-2}$), indicating that inherent numerical noise within each simulation is the dominant source of variability. At the resolution used, the simulations are stable with respect to run-to-run divergence, and the observed variations are primarily due to stochastic noise and shot noise. In practical terms, this means that repeated runs converge toward an expected mean behavior, rather than representing distinct but equally valid realizations of the system. A higher systematic variation would instead imply that each simulation outcome could diverge meaningfully over time. Differentiating between these two regimes (noise-dominated or systematically divergent) is crucial for assessing the reproducibility of cosmological simulations. We refrain from quantifying such distinctions here, as this would require a more complete set of tests, but the framework serves as a potential tool for future studies aimed at assessing the limits of reproducibility and predictive reliability in cosmological simulations.

The results presented here are specific to the configuration and feedback implementation of the \textsc{OpenGadget3} simulation, emphasizing the need for careful consideration when extrapolating findings to other simulation setups. To strengthen the generalization of our conclusions, future studies will include resolution-based convergence tests to assess the stability of these variation patterns at higher resolutions. Mergers, an important source of variability in galaxy properties, were not included in this analysis but represent a key factor that will be addressed in subsequent research to provide a complete picture of variability in galaxy cluster simulations.

In conclusion, this study provides a benchmark for the expected magnitude of variation in simulations run with \textsc{OpenGadget3}. The results highlight the importance of considering repeated-run variability when interpreting simulation outcomes, especially for properties sensitive to highly non-linear processes like star formation, black hole dynamics and feedback. Understanding these variations is crucial for accurately comparing simulation results with observational data and improving the reliability of their predictions.

Future work will extend this analysis to higher-resolution simulations; additionally, the exploration of a broader range of subgrid models and feedback parameters will help clarify the conditions under which feedback mitigates or exacerbates variability, thereby enhancing our understanding and interpretation of results of galaxy evolution in cosmological simulations.

\section*{Acknowledgements}
We appreciate the anonymous referees for their helpful suggestions. We sincerely thank Francesco Pauli for helpful discussions and timely advice, which greatly contributed to clarifying statistical methods used in this study. 
This work has been supported by the Spoke-3 "FutureHPC \& BigData" of the ICSC – Centro Nazionale di Ricerca in High Performance Computing, Big Data and Quantum Computing – and hosting entity, funded by European Union – NextGenerationEU. Supported by Italian Research Center on High Performance Computing Big Data and Quantum Computing (ICSC), project funded by European Union - NextGenerationEU - and National Recovery and Resilience Plan (NRRP) - Mission 4 Component 2 within the activities of Spoke 2 (Fundamental Research and Space Economy), (CN 00000013 - CUP C53C22000350006). This work has also been supported by the National Recovery and Resilience Plan (NRRP), Mission 4, Component 2, Investment 1.1, Call for tender No. 1409 published on 14.9.2022 by the Italian Ministry of University and Research (MUR), funded by the European Union – NextGenerationEU– Project Title "Space-based cosmology with Euclid: the role of High-Performance Computing" – CUP J53D23019100001 - Grant Assignment Decree No. 962 adopted on 30/06/2023 by the Italian Ministry of Ministry of University and Research (MUR).
AR acknowledges EuroHPC Joint Undertaking for awarding the project ID EHPC-REG-2024R01-029  access to Leonardo at CINECA, Italy.
AR acknowledges ISCRA for awarding this project access to the LEONARDO supercomputer, owned by the EuroHPC Joint Undertaking, hosted by CINECA, Italy (HP10BUFI59).
This work is also supported by  "Bando per il finanziamento della Ricerca Fondamentale 2024 dell’Istituto Nazionale di Astrofisica" Decreto n. 8/2024 MiniGrant "Numerical Chaos in Cosmological Simulations".
MV also acknowledges partial financial support from the INFN Indark Grant.

\appendix

\section{Mechanisms Contributing to Variability and Sensitivity in Cosmological Simulations}
\label{sec:appendix_variability_sources}

The sensitivity of cosmological simulations to divergent outcomes arises from several distinct mechanisms. While some represent fundamental numerical uncertainties (such as discretization errors, precision errors), others (such as subgrid physics models) act as triggers that amplify small perturbations into macro-scale variations.

We emphasize a distinction established in previous studies regarding bit-wise reproducibility: while "numerical determinism" can be enforced via specific implementation strategies (such as strict operation ordering or 128-bit summation with subsequent truncation), these methods impose significant computational overhead. Additionally, for complex subgrid models such as MUPPI \citep{murante_subresolution_2010, murante_simulating_2014}, which utilize Runge-Kutta integration with adaptive time steps, achieving such reproducibility is technically difficult and often impractical compared to simpler threshold-based subgrid schemes.
Ultimately, such determinism is primarily a tool for debugging and code verification. It does not imply that a physical solution is unique or insensitive to perturbations. It simply means that the same numerical approximations are reproduced across runs; however, when execution conditions differ (e.g. MPI task configuration, checkpoint–restart cycles, or hardware architecture), small perturbations can arise and may later be amplified by nonlinear dynamics. As demonstrated by \cite{genel_quantification_2019}, even when a code produces byte-identical results, the introduction of infinitesimal perturbations to the initial conditions leads to divergent galaxy properties.

This section establishes a systematic framework for understanding different mechanisms of simulation unpredictability with a particular focus on cosmological simulations (see e.g.  \cite{hernquist_discreteness_1993, binney_galactic_2008, barnes_error_1989}).

\paragraph{Poisson Noise from Finite Particle Sampling}
In particle-based simulations, continuous matter fields are represented by discrete particles, introducing statistical fluctuations in measured properties; commonly referred to as shot noise. This represents a fundamental precision limit that depends solely on resolution, not on the underlying physics.

\paragraph{Numerical Method and Precision Effects}
\begin{enumerate}
    \item {Discretization Errors:} Approximating continuous equations with discrete time steps or spatial grids introduces errors that depend on resolution choices. Large time steps may fail to capture system dynamics accurately.
    \item {Round-off Errors:} Finite precision computer arithmetic accumulates small errors at each computational step, particularly problematic in long-term integrations of sensitive systems.
    \item {Truncation Errors:} Approximating infinite series or integrals by finite sums can introduce significant errors if higher-order terms are important but neglected.
    \item {Inadequate Resolution:} Insufficient spatial or temporal resolution can lead to unphysical results, such as failing to resolve shock waves in hydrodynamic simulations or missing important small-scale physics.
    \item {Parallel Execution and Operation Ordering}: Because floating-point arithmetic is non-associative, differences in the order of operations (e.g. in distributed computing, multi-threading, vectorization) can introduce small but measurable numerical differences between runs.
\end{enumerate}

\paragraph{N-body Specific Numerical Effects}
Gravitational simulations face additional technical challenges:
\begin{enumerate}
    \item {Force Softening:} The softening length $\epsilon$ prevents numerical divergences from close encounters but introduces a trade-off between stability and accuracy. Inappropriate choices can either over-smooth important physics or reintroduce numerical instabilities.
    \item {Truncation of Multipole Expansion:} Tree codes approximate distant forces using multipole expansions. While typically well-controlled, these approximations can contribute to accumulated errors.
    \item {Improper Boundary Conditions:} Incorrectly specified boundaries can introduce spurious artifacts that destabilize the simulation.
\end{enumerate}

\paragraph{Probabilistic Subgrid Implementations}
Modern cosmological simulations incorporate probabilistic algorithms to model physical processes below the resolution limit, such as star formation, stellar feedback, and black hole accretion. While these models are typically implemented using random number generators (RNGs) with fixed seeds to ensure numerical determinism, their formulation involves probabilistic decision criteria. For example, small numerical perturbations can shift a gas particle's state across a star formation threshold, leading to differences in the timing or ordering of star formation events between realizations. In nonlinear regimes, such differences can be amplified, allowing small-scale perturbations to manifest as macroscopic variations between otherwise identical runs.

\paragraph{Event-Driven Amplification}
Discrete nonlinear events (such as galaxy mergers, major feedback episodes, or dynamical instabilities) can amplify small pre-existing differences into large, persistent variations. These effects produce step-function changes in system evolution that persist over long timescales \cite{keller_chaos_2019}.

\paragraph{Chaos}
Chaotic behavior refers to the sensitivity of a deterministic system to small perturbations in its initial conditions, which can propagate and lead to complex, large-scale changes in its evolution (Studies of chaos was pioneered by \cite{lorenz_deterministic_1963, mandelbrot_fractal_1982, may_simple_1976, feigenbaum_quantitative_1978}). This phenomenon, termed deterministic chaos, is characterized by positive Lyapunov exponents \citep{strogatz_nonlinear_2019}. Numerical simulations are inherently chaotic systems, and their long-term predictability is fundamentally limited by the amplification of small perturbations \citep{aarseth_dynamical_1963, miller_irreversibility_1964, standish_numerical_1968, henon_applicability_1964}. It was found that the Lyapunov timescale (\( \lambda^{-1} \)) generally decreases with increasing \( N \), but the trend saturates for \( N \gtrsim 30 \), indicating a regime where the divergence rate becomes largely independent of particle number (The question of how chaotic behavior scales with \( N \), referred to as "N-dependence", see some early works for example, \cite{gurzadian_problem_1984, gurzadian_collective_1986, heggie_chaos_1991, kandrup_time_1989, kandrup_how_1990, kandrup_divergence_1990}).

\subsection{Variability Studies in Cosmological Simulation}
\label{subsec:literature_review}
Systematic quantification of variability is not yet standard practice when reporting simulation results; examples of studies that have explicitly focused on quantifying or establishing baselines for such effects include \cite{su_feedback_2017,su_discrete_2018,oh_calibration_2020, granato_dust_2021, keller_uncertainties_2022, valentini_impact_2023}. Studying variability is particularly important in cosmological simulations, as it arises from the complex nonlinear interplay of gravity, hydrodynamics, and probabilistic subgrid physics such as star formation and feedback. These processes, together with mergers and environmental influences, can amplify small perturbations and lead to divergent galaxy properties across otherwise identical runs.

\cite{davies_quenching_2021,davies_galaxy_2022,davies_are_2023} examined the role of assembly history on galactic evolution using the EAGLE model \cite{schaye_eagle_2015}. They used a technique called genetic modification \citep{roth_genetically_2016,rey_quadratic_2018, hoffman_constrained_1991} to create different assembly histories for the galaxies. They found that mergers can significantly affect black hole (BH) growth and feedback, which in turn impacts the circumgalactic medium (CGM). The 2022 and 2023 follow-up studies expanded on the earlier work by focusing on the dynamic interaction between galaxy mergers, BH activity, and the CGM. Through more detailed simulations, they revealed how the timing and intensity of mergers influence the thermodynamic properties of the CGM (resulting in order-of-magnitude variations between their simulations) and the subsequent evolutionary path of the galaxy. The results suggest that understanding the subtleties of merger histories is crucial for accurately modeling galaxy formation and evolution, especially in relation to BH feedback and its long-term effects on galaxy properties.

One of the earliest systematic studies of variability within the context of cosmological simulations was conducted by \cite{thiebaut_onset_2008}, who demonstrated that in $\Lambda$CDM cosmologies, chaos arises predominantly at nonlinear scales. Using simulations of $100\,h^{-1}\,\text{Mpc}$ boxes with perturbed initial particle positions and velocities, they quantified the critical transition scale for simulation variability to be around $ 3.5\,h^{-1}\,\text{Mpc}$. On these scales, integrative quantities such as halo mass remained robust, while properties like spin orientation exhibited sensitivity to initial conditions.

Building on these ideas, \cite{genel_quantification_2019} performed analysis of variability in cosmological volumes ($25$–$50\,\text{Mpc}^3/h$) using the AREPO code and the full IllustrisTNG feedback model. Their resolution-scaled simulations demonstrated that tiny perturbations ($\approx 10^{-7}$ pc) in initial positions result in divergent galaxy properties of 2–25\% over cosmic time. They established that in simulations with full feedback, variability persists even at high resolution, whereas in no-feedback runs, variations reduce with resolution. Additionally, they conclude that the repeated-run divergence contributes significantly to the overall scatter in key scaling relations such as the Tully-Fisher and star formation rate–mass relations.

\cite{keller_chaos_2019} quantified variability in simulated galaxies by analyzing 128 identical simulations using GASOLINE2 and RAMSES codes. In Milky Way-like galaxies with $M_{\rm halo} \approx 6.5 \times 10^{11}\,M_\odot$, they found a typical scatter in stellar mass $\sigma_{M_*} \approx 5\%$ due to feedback-regulated star formation. Importantly, they noted that in the absence of feedback or under unregulated star formation, variations grow substantially, especially when gas is not rapidly exhausted. Additionally, the timing of mergers was shown to cause stellar mass variations by factors of 2, persisting for over a Gyr, thus linking such divergence to long-term galactic histories.

\cite{borrow_impact_2023} performed analysis with 16 identical simulations using the SWIFT code with cosmological volumes ($25 \text{ Mpc}^3$), finding stellar mass variations up to 25\%. Their results confirmed that while population-wide scaling relations remain robust, individual galaxy evolution, especially during "bursty" feedback or black hole accretion, can diverge significantly. Notably, they illustrated how random variability causes the correlation between galaxy stellar mass and black hole mass to essentially disappear at the highest black hole masses ($\text{M}_{\text{BH}} > 10^8 \text{M}_{\odot}$), underscoring the influence of non-linear feedback on black hole and galaxy co-evolution.

\cite{pakmor_quantifying_2025} investigated the variability of the Auriga galaxy formation model using seven Milky Way–like zoom-in realizations, differing only by random seeds. They found most global properties at $z=0$ (e.g. stellar mass, disc mass, star formation history) to be robust within $\leq10\%$, while instantaneous quantities such as star formation rate and internal disc structure showed larger variations (up to a factor of two). Additionally, they found that although satellite infall properties were consistent across runs, their subsequent orbital evolution diverged after the first pericentric passage. Importantly, the study further showed that changes in numerical resolution (by a factor of eight) introduced far greater and more systematic differences than the intrinsic stochastic variability.

These studies collectively demonstrate that cosmological simulations exhibit substantial variability arising from multiple, often coupled, sources. Isolating the contribution of each source individually is frequently infeasible and beyond the scope of most production simulation setups. Accordingly, \ref{sec:appendix_variability_sources} is intended to provide context for the inherent numerical and physical complexity of cosmological simulations, motivating our analysis of the cumulative impact of these factors on overall variability in simulation results.

\section{Variability in DMO and Hydro-only Simulations}
\label{sec:appendix_dmo}

\begin{figure*}[htbp]
    \centering
    \includegraphics[width=0.49\textwidth]{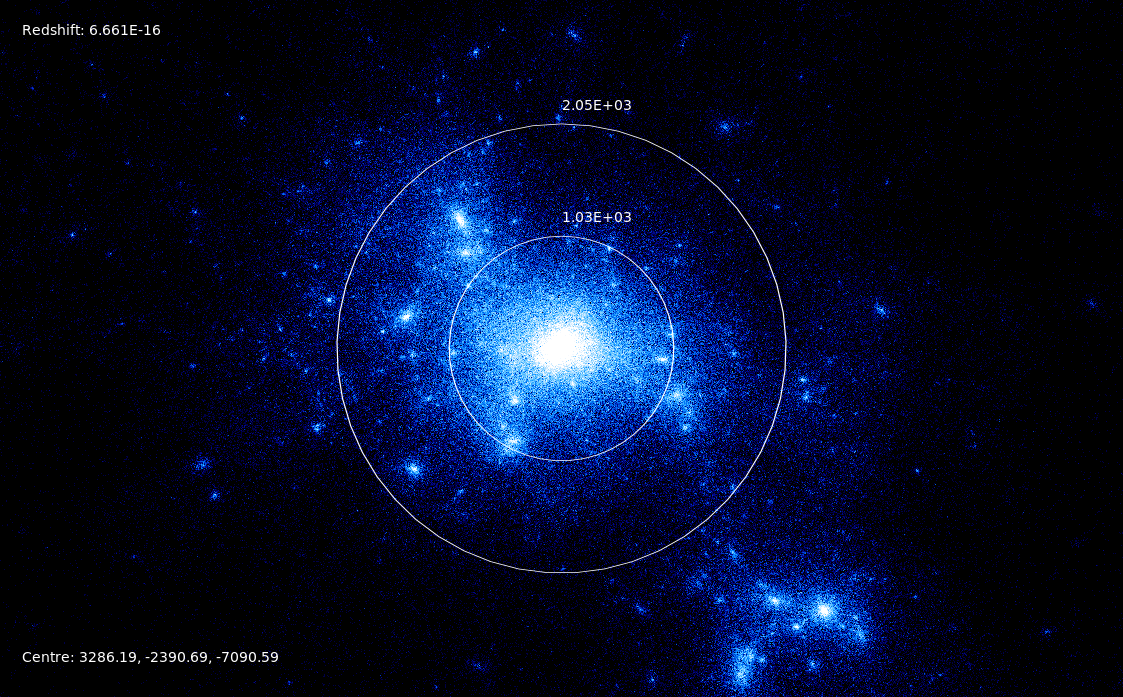}
    \includegraphics[width=0.49\textwidth]{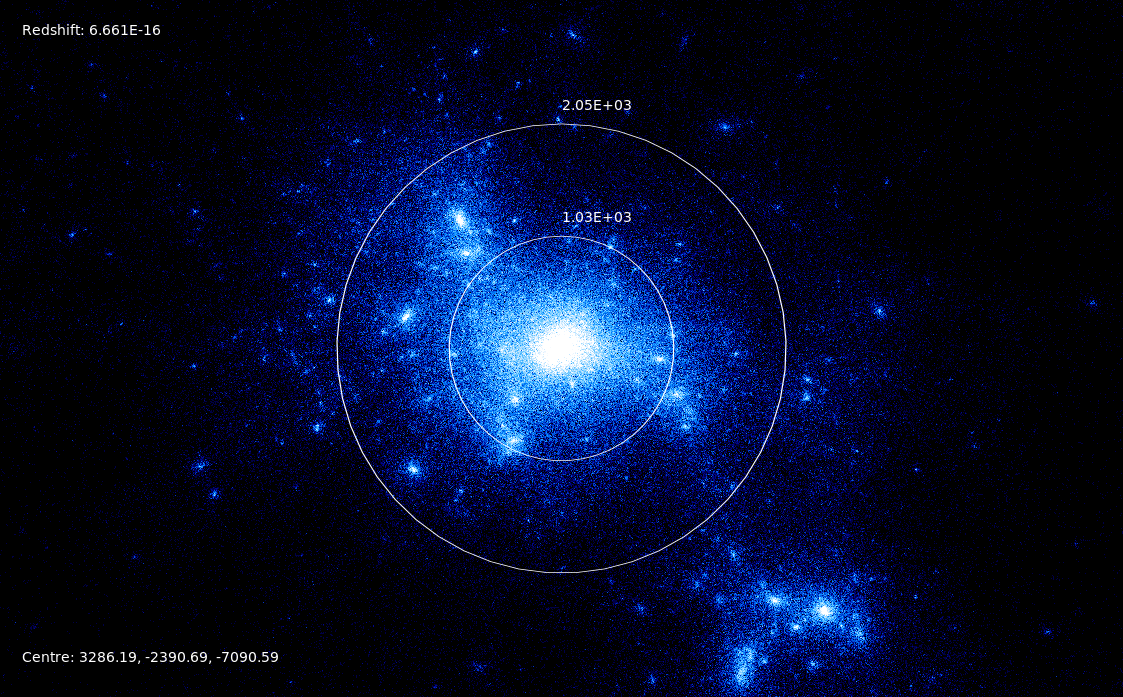}
    \includegraphics[width=0.49\textwidth]{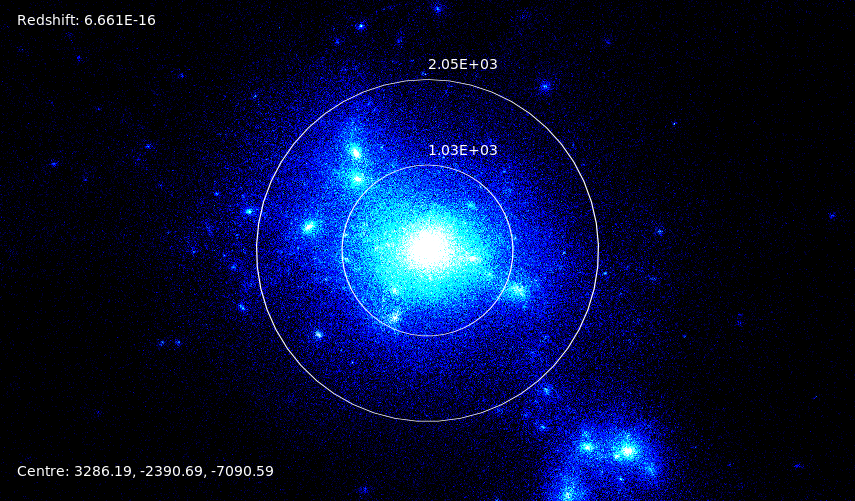}
    \includegraphics[width=0.49\textwidth]{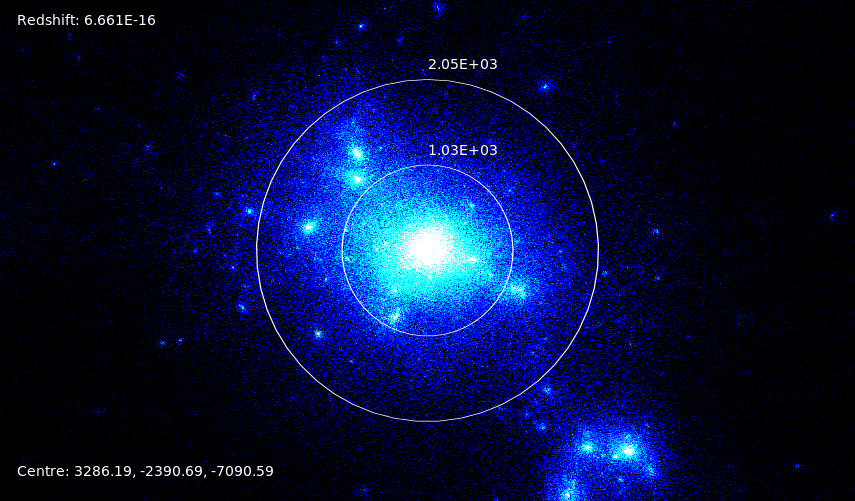}
    \caption{Visual comparison of the Dianoga D1 (see Table \ref{tab:simconfig} and Section \ref{sec:methodology}) dark-matter-only (top panels) and dark-matter and hydrodynamical simulations (bottom panels) at $z=0$, each with two realizations evolved from nearly identical set-up. The left panels shows the runs executed with 144 MPI tasks, while the right panels shows the run executed with 180 MPI tasks. No visually discernible differences are observed in the positions or morphology of the halo and its substructures in the respective repeated runs for dark-matter-only and as well as dark-matter and hydrodynamical simulations}
    \label{fig:dmohydrovisual}
\end{figure*}

\begin{figure*}[t]
    \centering
    \begin{minipage}[t]{0.45\textwidth}
        \centering
        \includegraphics[width=\linewidth]{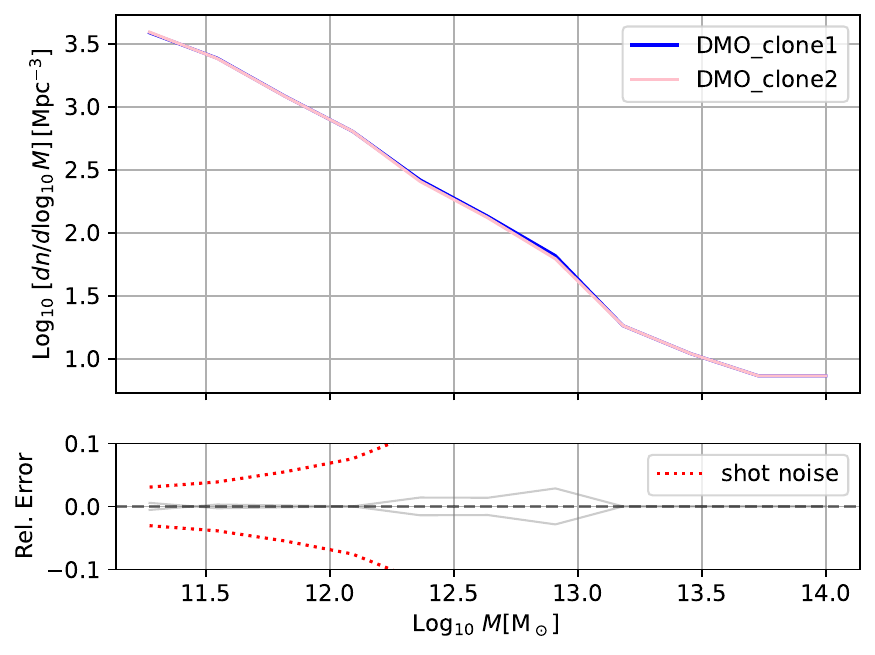}
    \end{minipage}
    \hfill
    \begin{minipage}[t]{0.45\textwidth}
        \centering
        \includegraphics[width=\linewidth]{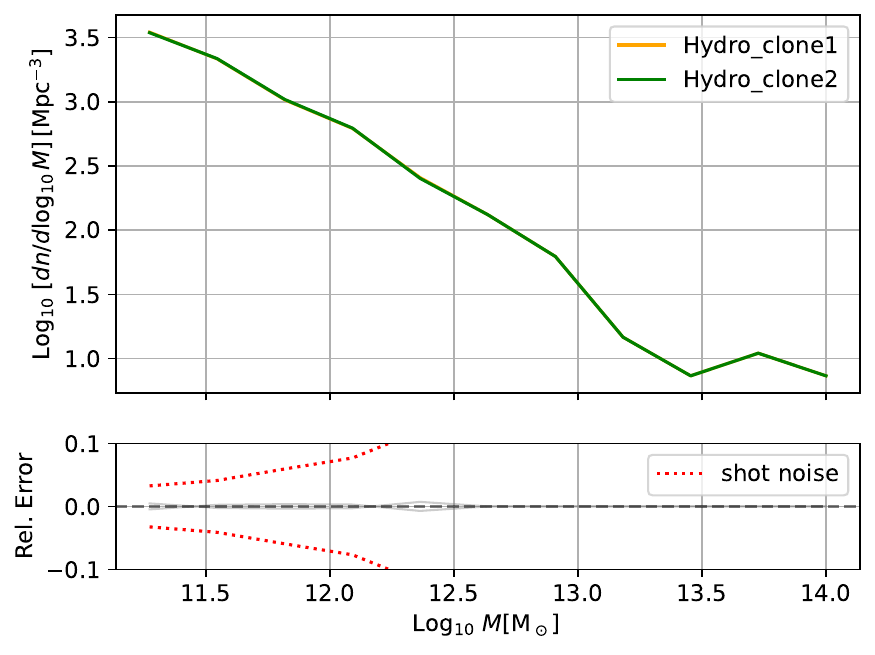}
    \end{minipage}
    \caption{Top panel shows Halo mass functions (HMFs) for the dark-matter-only (left panel, prefixed "DMO") and dark-matter and hydrodynamical simulations (right panel, prefixed "Hydro"). Each test set is comprised of 2 identical runs, executed with 144 MPI tasks (DMO\_clone1 and Hydro\_clone1) and 180 MPI tasks (DMO\_clone2 and Hydro\_clone2). The bottom panel shows the relative difference between the respective identical-realizations (gray) and the shot noise (red dotted line), demonstrating statistical agreement across the resolved mass range.}
    \label{fig:r_haloMassFunction_dmohydro}
\end{figure*}

In this section, we demonstrate the impact of intrinsic numerical variability by excluding the stochasticity associated with the subgrid physics used in our full-physics runs. We performed controlled tests for Dark Matter Only (DMO) and Dark Matter plus hydrodynamics (Hydro) simulations using our low-resolution setup (see Table \ref{tab:simconfig}) with a mixed precision configuration. For each test set, we evolved three realizations using identical initial conditions and numerical parameters. Two runs executed with identical MPI task counts (144 tasks) yielded byte-identical results.
A third run was evolved using a different domain decomposition (180 MPI tasks), introducing small perturbations in the ordering of floating-point operations relative to the 144-task run. Figure \ref{fig:dmohydrovisual} provides a visual comparison of the most massive halo at $z=0$ between the 144-task and 180-task runs for both DMO (top panel) and Hydro (bottom panel) cases. In contrast to the full-physics simulations presented in the main text (Figure 1), these realizations exhibit no visually discernible differences in their substructure. Furthermore, Figure \ref{fig:r_haloMassFunction_dmohydro} shows the Halo Mass Functions (HMFs) for the DMO (left panel) and Hydro (right panel) sets, with the relative difference between realizations provided in the lower panels. Both HMFs are statistically indistinguishable across the entire mass range, demonstrating that these outcomes are highly consistent and show low sensitivity to round-off level perturbations.

\section{Tests with Varied MPI Task Counts}
\label{sec:appendix_variedMPI}

\begin{figure*}[t]
    \centering
    \begin{minipage}[t]{0.45\textwidth}
        \centering
        \includegraphics[width=\linewidth]{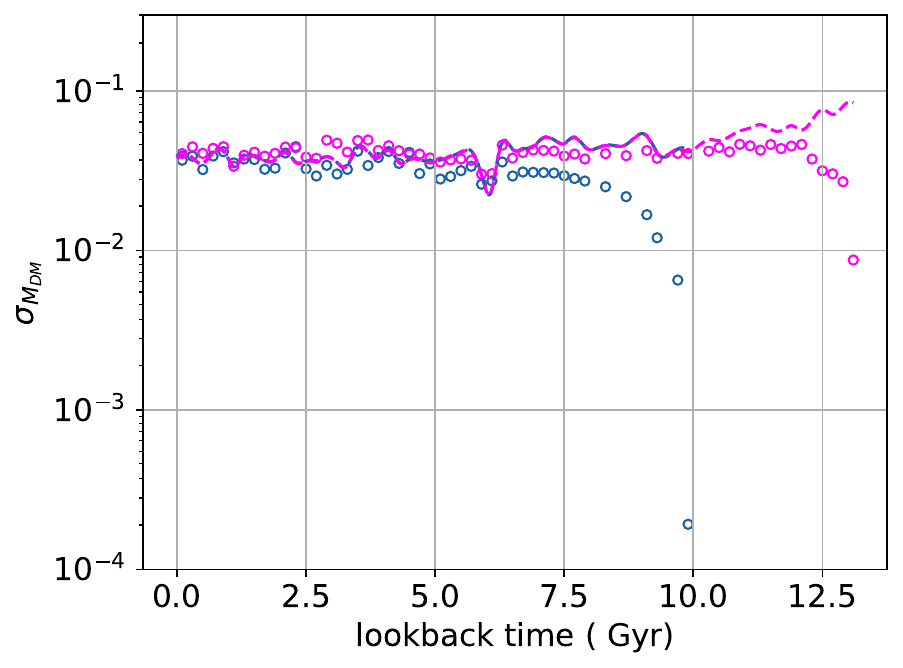}
    \end{minipage}
    \hfill
    \begin{minipage}[t]{0.45\textwidth}
        \centering
        \includegraphics[width=\linewidth]{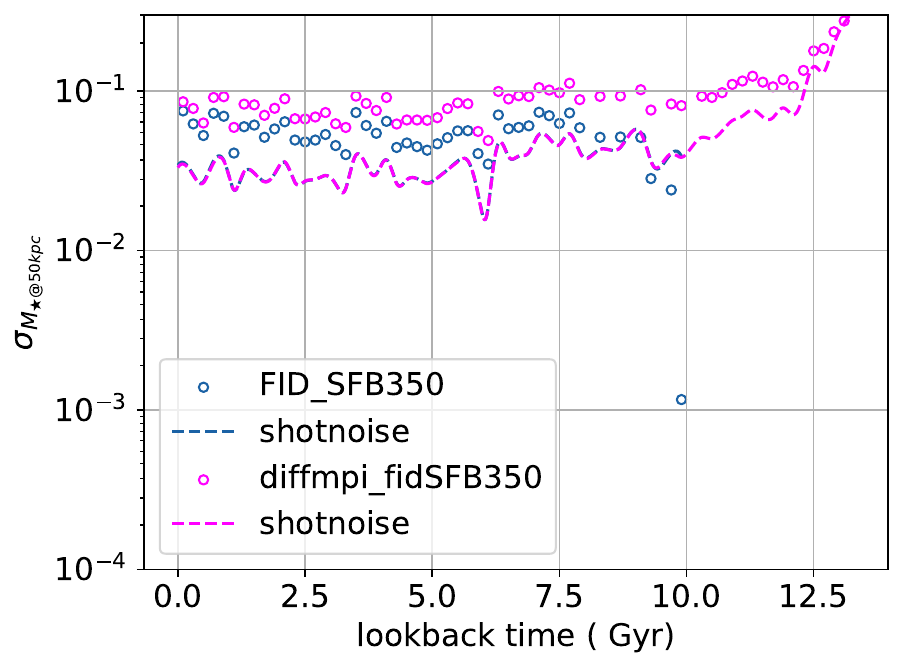}
    \end{minipage}
    \caption{Total variation (Method 1) in galaxy dark matter mass (left panel) and stellar mass (right panel) as a function of lookback time (Gyr), shown for the Fiducial-NOBH model. The blue lines represent the standard Fiducial set (fixed MPI tasks, as in Figure \ref{fig:FID_SFB350_mdm_mstar_mcold_50kpc_reg}), while the pink lines represent the varied MPI task set (label prefixed with 'diffmpi\_fid'). Dashed colored lines indicate the approximate shot noise floor for each test set.}
    \label{fig:FIDvsPERT_SFB350_mdmANDmstar_reg}
\end{figure*}
To assess the sensitivity of the results to small numerical perturbations at the outset, we repeated the Fiducial simulations (excluding black holes) with different MPI task decompositions. This set comprised four runs, executed with 144, 180, 216, and 252 MPI tasks, respectively. In Figure \ref{fig:FIDvsPERT_SFB350_mdmANDmstar_reg}, 
\(\sigma\) (computed with Method 1) for the varied-MPI-task Fiducial set is compared to the baseline Fiducial set performed with identical MPI task counts.

For $\sigma_{M_*}$ (right panel), the varied-MPI-task-runs exhibit an earlier onset of measurable variation, beginning at lookback times \(>12.5 Gyr\) , and $\sigma_{M_*}$ $>$ 0.2  that coincides with the higher shot noise and increased stochastic noise which can be attributed to peak of star formation. The variation then decreases and subsequently plateaus, approaching a level similar to the unperturbed baseline but remaining slightly higher (the difference likely reflects additional noise introduced by the altered MPI decomposition).

In contrast, the variation in dark matter mass (left panel of Figure \ref{fig:FIDvsPERT_SFB350_mdmANDmstar_reg}) shows no marked difference between the perturbed and unperturbed runs: though the variations in the varied-MPI-task set accumulate from \(>12.5\) Gyr, both plateau at nearly identical levels. The lack of sensitivity to small numerical perturbations reflects the inherently stable nature of collisionless dark matter dynamics.
For stellar mass, however, excess round-off error can act as an additional noise contribution, which may compound with the stochasticity of the star formation and feedback subgrid models.

\section{Variability including black holes: Galaxy cold gas mass and black hole accretion rate}
\label{sec:appendix_withbh_sfrbhmd}
\begin{figure*}[t]
    \centering
    \begin{minipage}[t]{0.45\textwidth}
        \centering
        \includegraphics[width=\linewidth]{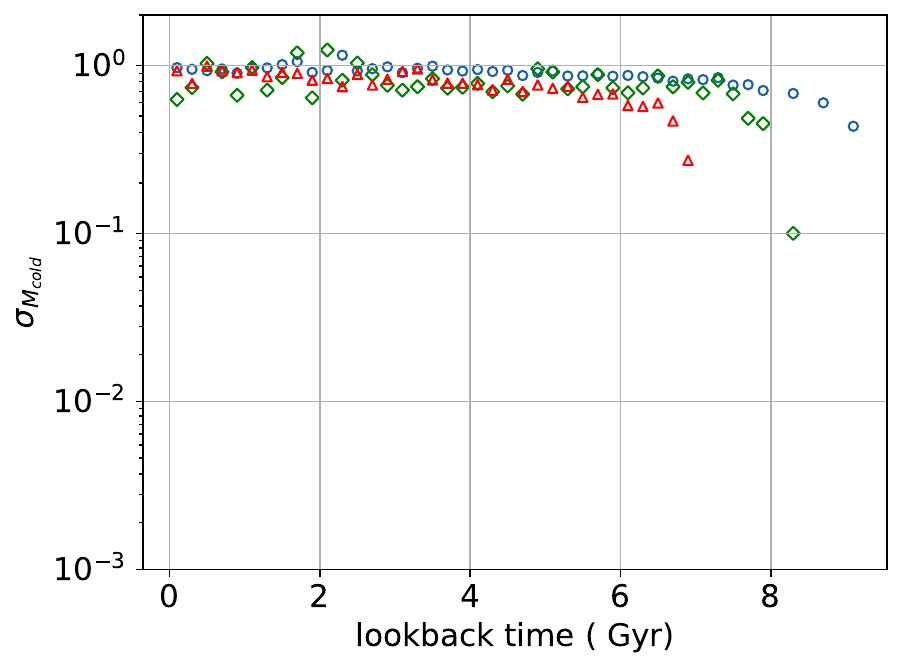}
    \end{minipage}
    \hfill
    \begin{minipage}[t]{0.45\textwidth}
        \centering
        \includegraphics[width=\linewidth]{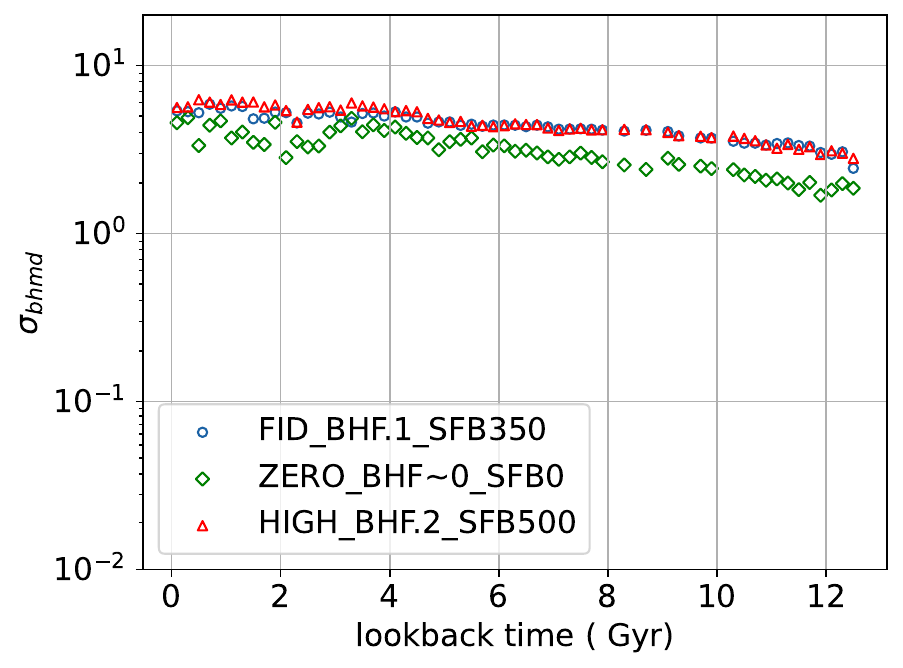}
    \end{minipage}
    \caption{Same as Figure \ref{fig:fbtest_D1_1x__mstar_50kpc_AndD1_1x__snapmbh_} but for galaxy cold gas mass (left panel) and black hole accretion rate (right panel).}
    \label{fig:fbtest_D1_1x__mcold_50kpc_AndD1_1x__bhmd_}
\end{figure*}
We show here the total variation (Method 1) in galaxy cold gas mass (\ref{fig:fbtest_D1_1x__mcold_50kpc_AndD1_1x__bhmd_} left panel) and black hole accretion rate (BHMD, \ref{fig:fbtest_D1_1x__mcold_50kpc_AndD1_1x__bhmd_} right panel) for the Fiducial, ZERO and HIGH feedback cases.

The variation in galaxy cold gas mass is broadly similar across all feedback configurations, plateauing at \(\sigma_{M_{\text{cold}}} \approx 1\). For BHMD, though variation in black hole mass accretion exhibits a slightly lower scatter for the ZERO-feedback case, likely due to the presence of more massive black holes that largely depleted their gas reservoirs; overall, \(\sigma_{\text{bhmd}} \approx 4\text{–}6\) across all cases. These variations demonstrate that black hole growth remains a highly stochastic process even when feedback strength varies. Note that these quantities are highly variable because they are instantaneous rather than time-averaged and are strongly influenced by short-term environmental changes.

\section{Mixed precision variability: Galaxy dark matter and black hole mass}\label{sec:appendix_mpdp_mdmmbh}
\begin{figure*}[t]
    \centering
    \begin{minipage}[t]{0.45\textwidth}
        \centering
        \includegraphics[width=\linewidth]{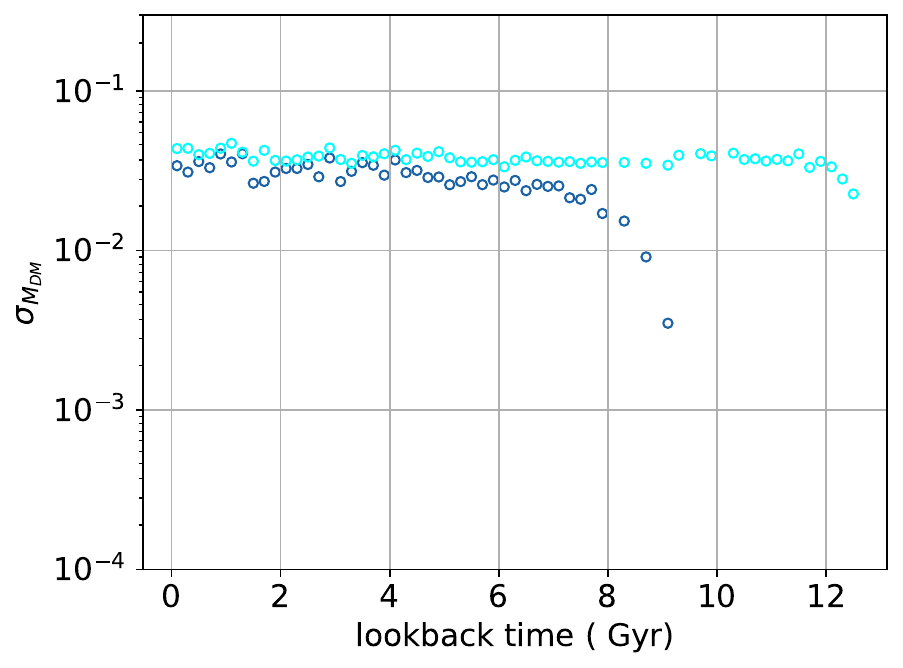}
    \end{minipage}
    \hfill
    \begin{minipage}[t]{0.45\textwidth}
        \centering
        \includegraphics[width=\linewidth]{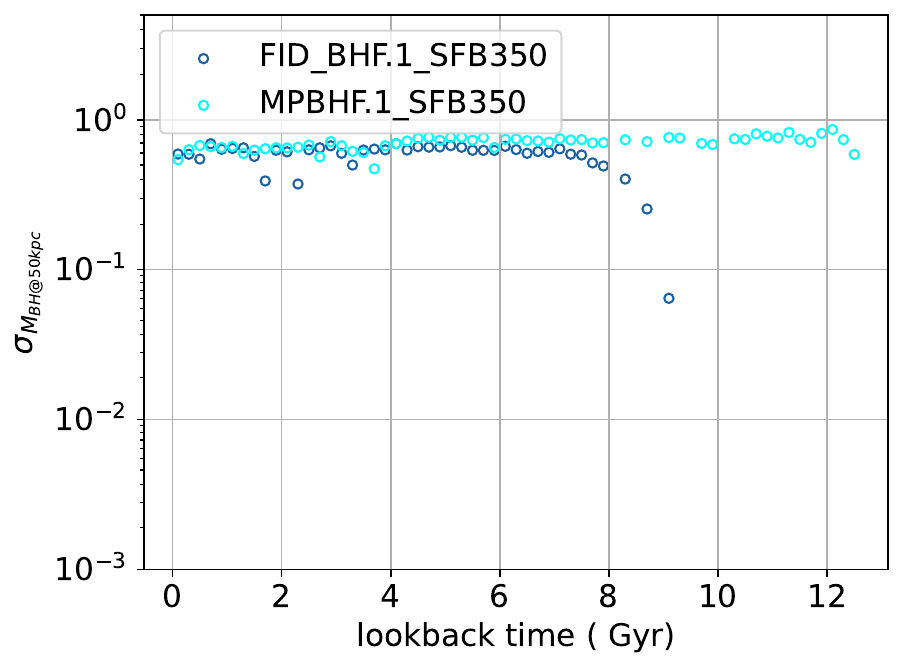}
    \end{minipage}
    \caption{Total variation (Method 1) in galaxy dark matter mass (left panel) and black hole mass (right panel); as a function of lookback time (Gyr), shown for Fiducial double (in blue) and mixed precision sets (MP, in cyan).}
    \label{fig:D1_1x__mdm_MP_MBH}
\end{figure*}

This Appendix provides supplementary figures in support of Section \ref{mixedprecision}. Figure \ref{fig:D1_1x__mdm_MP_MBH} displays the total variation ($\sigma$) computed via Method 1 for galaxy dark matter mass (left panel) and black hole mass (right panel) across both the double and mixed precision test sets. While the accumulation of round-off errors occurs more rapidly in the mixed precision runs (well before $ 9\,\mathrm{Gyr}$), the resulting variability in these specific properties is not significantly varied. Both $\sigma_{M_{\text{dm}}}$ and $\sigma_{M_{\text{mbh}}}$ plateau at levels nearly identical to their double precision counterparts, demonstrating minimal sensitivity to numerical precision effects.

\section{Subgrid Physics Implementation Details}\label{sec:appendix_intro_sf}
\subsection{Star formation and associated feedback} 
Star formation and stellar feedback are incorporated through a set of subgrid models that convert gas into stars and model the impact of stellar feedback on the surrounding gas. For the reference model of star formation and feedback implemented, please check \cite{springel_cosmological_2003, tornatore_cooling_2003, tornatore_chemical_2007}. Here we only summarise some implementation details.

The Interstellar Medium (ISM) refers to the diffuse matter i.e. gas (atomic, molecular, and ionized) and dust filling the space between stars within a galaxy. The ISM plays a critical role in star formation, as stars form from the gravitational collapse of dense regions within the ISM. This aspect is modeled with gas particles, where a given gas particle is defined to be multiphase with respect to a certain density threshold and temperature threshold (which defines a hot or cold phase). Thus,
\begin{itemize}
    \item A density threshold: 
   A gas particle can form stars if its density exceeds a threshold value \( \rho_{\text{th}} \). This threshold is chosen from Schmidt law-like prescription and derived from density function sustaining a self-regulated star formation regime.
   \item A temperature criterion: 
   Gas particles eligible for star formation must also be sufficiently cool, typically below a temperature threshold \( T_{\text{th}} \). This ensures that only cold, dense gas, which is capable of collapsing under its own gravity, forms stars.
\end{itemize}

\paragraph{Star Formation Rate (SFR):} 
quantifies the rate at which a galaxy forms new stars over a given period of time, typically expressed in solar masses per year (\( M_\odot \, \text{yr}^{-1} \)). In GADGET-3, once the particle is converted to a star particle it follows a Schmidt law-like prescription, where the rate at which gas converts into stars is proportional to the gas density. The instantaneous SFR density for a gas particle \( i \) is given by:

\begin{equation}
\dot{\rho}_{\star, i} =  (1-\beta) \frac{\rho_i}{t_{\star}}
\end{equation}

where:
\begin{itemize}
    \item  \( \dot{\rho}_{\star, i} \) is the star formation rate density,
    \item  \( \beta \) is the mass fraction of massive stars (dependent on the Initial Mass Function, IMF),
    \item  \( \rho_i \) is the gas density of particle \( i \),
    \item  \( t_{\star} = \frac{t_{\text{dyn}}}{c_{\star}} \) is the star formation timescale configured according to Schmidt's law \cite{schmidt_rate_1959}.
\end{itemize}

IMF used in this work is \cite{chabrier_galactic_2003}. The actual process of star formation is implemented stochastically, converting gas particles into star particles according to the computed SFR. 

\paragraph{Stellar Feedback}

There are a variety of stellar feedback models implemented in GADGET-3. Below, we focus only on  the ones presented by
\cite{springel_cosmological_2003, tornatore_chemical_2007}.
Stellar feedback in GADGET-3 encompasses thermal and kinetic feedback, including primarily supernova explosions, winds, and chemical enrichment. These processes inject energy, momentum, and heavy elements into the surrounding gas, influencing its thermal and dynamical state.
The thermal energy released is computed based on the IMF adopted. The feedback heats up the hot gas as well as evaporates the gas in the cold phase via thermal conduction. This recycles in the dense gas clouds back to the ambient phase.
The evaporated cold gas mass follows,
   \begin{equation}
   \dot{\rho}_{c} = -A\beta \frac{\rho_{c}}{t_{\star}}
   \end{equation}
where, A is the evaporation efficiency and assumed to follow a theoretically modeled dependence on the ISM density (\(A \propto \rho^{-\frac{4}{5}} \))

Radiative cooling of the hot gas replenishes the cold gas clouds. The cycle continues sustaining a self-regulated process of star formation.

Instead of explicitly modeling the detailed mass exchange between phases (cold clouds, hot gas, and stars), the method assumes equilibrium conditions are achieved rapidly. 
The model evolves gas particles probabilistically into star particles based on the computed star formation rate. For each timestep, a new star particle is spawned if a random number drawn uniformly falls below the probability:
\begin{equation}
p_\star = \frac{m}{m_\star} \left( 1 - \exp\left[ - \frac{(1 - \beta)x \Delta t}{t_\star} \right] \right),
\end{equation}
where \( m_\star \) is the mass of a single star particle, and $x$ is the fraction of gas in cold clouds.

\label{intro_sfwinds}In the multiphase model for ISM supporting SF and feedback, winds are an additional implementation following the phenomenological prescription for galactic winds, also, as detailed in \cite{springel_cosmological_2003}.
Hot gas associated with supernovae is converted into wind particles following a mass loss proportional to the SFR  : 
\begin{equation}
   \dot{M}_{w} = \eta \dot{M}_{\star} 
   \end{equation}

   where:
   $\eta$ is the efficiency parameter
   (typically \(\eta\) =2 is assumed). This conversion depends on probability of a generated random number to be less than a limit of:
\begin{equation}p_w = 1- \exp\left[-\frac{\eta(1-\beta)x \Delta t}{t_{\star}}\right]\end{equation}

where:
\begin{itemize}
    \item \( p_w \) is the probability that a gas particle is converted into a wind particle within the time interval \(\Delta t\).
    \item \( \Delta t \) is the simulation time step.
    \item \( t_{\star} \) is the star formation timescale.
\end{itemize}

Depending on the gradient of the gravitational potential of the galaxy, the wind particles are ejected orthogonally, along the rotational axis of the galaxy. The velocity of the winds are configured, a \(v_w\) = 350 km \(s^{-1}\) corresponds to a fixed fraction of 50\% of the supernovae energy converted into the kinetic energy carried by the wind \citep{bassini_buchi_2021}. 

Supernovae feedback is implemented together with metal enrichment following the prescriptions presented in \cite{tornatore_chemical_2007}. A metal dependent cooling function is used to compute the respective energy radiated due to the quantity of metals produced.   
H and the by products  He, C, Ca, O, N, Ne, Mg, S, Si, Fe, Na, Al, Ar  are all followed and computed specific to the IMF and predictions from theoretical modeling respective to type-Ia, II and asymptotic giant branch (AGB) stars; see \cite{matteucci_chemical_1993}, \cite{thielemann_supernova_2003}, \cite{woosley_evolution_1995}, \cite{romano_quantifying_2010} and \cite{karakas_vizier_2010}.

\subsection{Black hole and associated feedback} \label{sec:appendix_intro_blackhole}
The base reference for black hole implementation is \cite{springel_modelling_2005}.
Black holes are initially seeded in dark matter halos once they reach a critical mass or are otherwise designated to host a BH. Typically, seed BHs have masses around \(10^5 - 10^6 M_{\odot}\), placed in the center of newly formed halos. 

The accretion rate onto a black hole is determined using the Bondi-Hoyle-Lyttleton formula (refer \cite{bondi_spherically_1952}), modified for a realistic astrophysical context:

\begin{equation}
\dot{M}_{\text{BH}} = \alpha \frac{4 \pi G^2 M_{\text{BH}}^2 \rho}{(c_s^2 + v^2)^{3/2}}
\end{equation}

where:
\begin{itemize}
    \item \( \dot{M}_{\text{BH}} \) is the accretion rate onto the black hole,
    \item \( \alpha \) is a dimensionless parameter that accounts for the efficiency of the accretion process,
    \item \( G \) is the gravitational constant,
    \item \( M_{\text{BH}} \) is the black hole mass,
    \item \( \rho \) is the local gas density,
    \item \( c_s \) is the sound speed of the gas,
    \item \( v \) is the relative velocity between the black hole and the gas.
\end{itemize}

This accretion rate is capped by the Eddington limit to avoid super-Eddington accretion scenarios:

\begin{equation}
\dot{M}_{\text{Edd}} = \frac{4 \pi G M_{\text{BH}} m_p}{\epsilon_r \sigma_T c}
\end{equation}

where:
\begin{itemize}
    \item \( \dot{M}_{\text{Edd}} \) is the Eddington accretion rate,
    \item \( m_p \) is the proton mass,
    \item \( \epsilon_r \) is the radiative efficiency,
    \item \( \sigma_T \) is the Thomson scattering cross-section,
    \item \( c \) is the speed of light.
\end{itemize}

\label{intro_blackhole_fb}\paragraph{Feedback Mechanisms:}The energy feedback from accreting black holes is modeled as thermal energy injected into the surrounding gas. The rate of energy injection is given by:

\begin{equation}
\dot{E}_{\text{feed}} = \epsilon_f \epsilon_r \dot{M}_{\text{BH}} c^2
\end{equation}

where:
\begin{itemize}
    \item \( \dot{E}_{\text{feed}} \) is the feedback energy rate,
    \item \( \epsilon_f \) is the fraction of the radiated energy that couples to the ISM,
    \item \( \epsilon_r \) is the radiative efficiency,
    \item \( \dot{M}_{\text{BH}} \) is the black hole accretion rate.
\end{itemize}

This energy transfers to the closest 200 of gas particles, increasing the pressure and temperature, which can suppress further star formation. \( \epsilon_f \) and \( \epsilon_r \) are configured to match the empirical  \( M_{\text{BH}} \)- \( M_{\star} \) correlation obtained from star-forming regions in galaxies

Local gas properties such as density, temperature, and bulk velocity are calculated using the SPH smoothing kernel (similar to SPH particles), providing the necessary inputs for computing the accretion rate using the Bondi-Hoyle-Lyttleton formula. Gas particles within the smoothing kernel of a black hole are assigned a probability \( p_j \) of being accreted:
\begin{equation}
p_j = w_j \frac{\dot{M}_{\text{BH}} \Delta t}{\rho},
\end{equation}
where:
\begin{itemize}
    \item \( w_j \) is the SPH kernel weight,
    \item \( \dot{M}_{\text{BH}} \) is the accretion rate,
    \item \( \Delta t \) is the timestep, and
    \item \( \rho \) is the local gas density.
\end{itemize}
A random number determines whether a particle is accreted. 

Apart from accretion, the rate of feedback is also computed from a given black hole. Feedback energy proportional to the accreted mass-energy (\( \dot{E}_{\text{feed}} = \epsilon_f \epsilon_r \dot{M}_{\text{BH}} c^2 \)) is deposited into surrounding gas particles, weighted by the SPH kernel.

\paragraph{Black Hole Dynamics:}
Black holes are treated as sink particles that interact gravitationally with their environment. Their mass grows through accretion and mergers with other black holes. The dynamics of BH particles is followed through the simulation using standard N-body techniques, ensuring accurate gravitational interactions.

\subsection{Details of the Random Number Generation}
The Random Number Generator (RNG) is used to model stochastic processes such as the probability of formation of star or wind particle (detailed in the above subsections \ref{sec:appendix_intro_sf} and \ref{sec:appendix_intro_blackhole}).
\textsc{OpenGadget3} utilizes the GNU Scientific Library (GSL), specifically the high-quality ranlxd1 algorithm (a variant of the high-quality, long-period pseudo-random number generator Ranlux algorithm \citep{marsaglia_new_1991}).
To guarantee parallel-safety and bit-wise reproducibility of the random number assignment across parallel executions, the code employs a deterministic lookup system. The global GSL RNG is initialized with a fixed global seed and used once at the start to populate a static lookup table.
When a particle requires a random number, the value is retrieved from this table using a deterministic index which is calculated based on the particle's unique ID.
This deterministic ID-based indexing ensures that the random number associated with a given particle and process is invariant across different parallel execution environments (MPI ranks, load balancing).

This implementation ensures numerical reproducibility for a fixed execution path; however, in other cases, small differences can cause RNG sequences across the entire volume to become uncorrelated between the two runs. An alternative approach in the SWIFT code \citep{schaller_swift_2024}, utilizes a combination of particle IDs and discrete timestamps to ensure more localized and time-consistent random sequences. While the relative contribution of the RNG scheme versus the inherent non-linearity of the subgrid physics remains an interesting open question for future benchmarking, our results establish a baseline for the variability present in the current widely-used framework. A possible future direction of investigation will be to include different approaches to RNG within \textsc{OpenGadget3}, so as to assess any impact that this can have on our simulation results.


\bibliographystyle{elsarticle-harv} 
\bibliography{references}






\end{document}